\documentclass[prd,aps,eqsecnum,floatfix,nofootinbib,preprint,tightenlines]{revtex4}

\usepackage{latexsym}
\usepackage{amsmath}
\usepackage{graphicx}

\def\bibi{\bibitem}

\let\ced=\c                     % cedilla
\def\a{\alpha}
\def\b{\beta}
\def\c{\chi}
\def\d{\delta}
\def\g{\gamma}

\def\k{\kappa}

\def\m{\mu}
\def\n{\nu}

\def\p{\pi}                     % Also, \varpi
\def\r{\rho}                    %       \varrho
\def\t{\tau}

\def\P{\Pi}

\def\cd{{\cal D}}

\def\cp{{\cal P}}

\def\cbo{{\,\raise-.15ex\Sc [\,}}                       % curly "
\def\ddt#1{{\buildrel {\hbox{\LARGE .\kern-2pt.}} \over {#1}}}% double dot-over
\def\ie{\mbox{\it i.e.}}
\def\eg{\mbox{\it e.g.}}
\def\etc{\mbox{\it etc.}}

\def\leqx{\,\raisebox{-1.0ex}{$\stackrel{\textstyle <}{\sim}$}\,}

\def\floatcaption#1#2{ \caption{ #2 \ [#1] \label{#1}} }
\def\floatcaption#1#2{ \caption{#2 \label{#1}} }

\def\bibi{\bibitem}    %% uncomment to suppresses citation labels

\def\ttl#1{{\it #1}}
\def\ttl#1{}

\long\def\symbolfootnote[#1]#2{\begingroup%
\def\thefootnote{\fnsymbol{footnote}}\footnote[#1]{#2}\endgroup}

\long \def \blockcomment #1\endcomment{}

\def\eg{{\it e.g.}}
\def\etc{{\it etc.}}
\def\seef{{\it cf.}}

\def\bu{\overline{u}}

\def\hw{\hat{w}}

\def\paper{{$\cp$1}}

\def\textit#1{{\it #1}\kern.1em }

\def\ansatz{{\it ansatz}}

\begin{document}
\rightline{TTK-12-06}

\begin{center}
\begin{boldmath}
{\large\bf An updated determination of  $\bf\a_s$ from $\bf\t$ decays}\\[0.4cm]
\end{boldmath}
\vspace{3ex}
{Diogo~Boito,$^a$ Maarten~Golterman,$^b$ Matthias~Jamin,$^c$ 
Andisheh~Mahdavi,$^b$ Kim~Maltman,$^{d,e}$ James~Osborne,$^b$ Santiago~Peris$^b$%
\symbolfootnote[2]{Permanent address: Department of Physics, Universitat Aut\`onoma de Barcelona, E-08193 Bellaterra, Barcelona, Spain}
\\[0.1cm]
{\it
\null$^a$Institut f\"ur Theoretische Teilchenphysik und Kosmologie\\
RWTH Aachen University, D-52056 Aachen, Germany\\
\null$^b$Department of Physics and Astronomy\\
San Francisco State University, San Francisco, CA 94132, USA\\
\null$^c$Instituci\'o Catalana de Recerca i Estudis Avan\ced{c}ats (ICREA)\\
IFAE,  Universitat Aut\`onoma de Barcelona\\ E-08193 Bellaterra, 
Barcelona, Spain\\
\null$^d$Department of Mathematics and Statistics\\ 
York University,  Toronto, ON Canada M3J~1P3\\
\null$^e$CSSM, University of Adelaide, Adelaide, SA~5005 Australia}}
\\[6mm]
{ABSTRACT}
\\[2mm]
\end{center}
\begin{quotation} 
Employing our previous framework to  treat
non-perturbative effects self-consistently, including duality violations, we update the
determination of the strong coupling, $\a_s$, using a modified version 
of the 1998 OPAL data, updated to reflect current values of exclusive 
mode hadronic $\tau$ decay branching fractions.
Our best $n_f=3$ values from the updated OPAL data are
$\a_s(m^2_\t)=0.325\pm 0.018$ and $\a_s(m^2_\t)=0.347\pm 0.025$ in
fixed-order and contour-improved perturbation theory, respectively.

To account for non-perturbative effects, non-linear, multi-parameter fits are
necessary. 
We have, therefore, investigated
the posterior probability distribution of the model parameters 
underlying our fits in more detail. We find that OPAL data alone 
provide only weak constraints on some of the parameters needed
to model duality violations, especially in the case of
fits involving axial vector channel data, making additional prior assumptions 
on the expected size of these parameters necessary at present.
We provide evidence that this situation could be greatly 
improved if hadronic spectral functions based on the high-statistics 
BaBar and Belle data were to be made available.

\end{quotation}

\vfill
\eject
\setcounter{footnote}{0}

%%%%%%%%%%%%%%%%%%%%%%%%%%%%%%%%%%%%%%%%%%%%%%%%%%%%%%%%%%%%%%
%%####%%
%\newpage
\section{\label{intro} Introduction}
%%####%%
In a previous article \cite{BCGJMOP},
 henceforth referred to as \paper, we developed a 
new framework for the determination of the strong
coupling, $\a_s(m_\t^2)$, from non-strange vector ($V$) and axial-vector ($A$)
 hadronic $\t$-decay data.  The new framework
starts from the usual finite-energy sum-rule (FESR) analysis, but improves this
approach in two ways with regard to the small, but quantitatively
significant non-perturbative corrections present in the theoretical 
representation of the FESR spectral integrals below the $\t$ mass. 
First, contributions from higher orders in the operator product expansion
(OPE) are taken into account self-consistently.  Second, in view of the 
fact that duality violations (DVs) are clearly present in the
experimental spectral distributions,
we use an explicit
parametrization of violations of quark-hadron duality in our fits. 
As explained in detail in \paper, these
two improvements are intricately connected:
estimates of the non-perturbative contribution 
to the sum rules with controlled errors cannot be obtained
without taking both of these effects into account.
 
Our framework was tested in \paper\ by applying it to data from the OPAL collaboration \cite{OPAL}.  We showed that fits to the data using this new
framework are indeed feasible in practice.   The resulting value for
$\a_s$ acquires larger errors than seen in previous extractions of $\a_s$
from hadronic $\t$ decays.    The most important reason for this is that, in 
order to take DVs into account, our fits necessarily contain more parameters, while we are limited to presently available data.  

More recent data are in principle available.  First, there are the ALEPH
data \cite{ALEPH98}, updated in 2005/08 \cite{ALEPH,Davieretal08}. 
Presently, use of the 2005/08 ALEPH data is questionable
 because correlations due to unfolding were inadvertently
omitted in the 2005/08 ALEPH update and hence from the publicly available
covariance matrices~\cite{TAU2010}.   Alternatively, more precise
spectral functions can in principle be extracted from BaBar or Belle data.
This would be very interesting, because one expects  such
spectral functions to have significantly smaller errors in the 
energy region near the $\t$ mass important for the extraction of $\a_s$.
We will argue in this article that it should be possible to determine 
the non-perturbative contributions to the sum rules, and thus $\a_s$, with much smaller 
errors were such data to become available.

It is nevertheless possible to make some progress with the OPAL data
beyond the results presented in \paper. The reason is that in 
Ref.~\cite{OPAL} the normalizations of the exclusive $\t$ decay modes,
as well as the values of a number of physical constants (such as the
$\t$ mass, the electronic branching fraction $B_e$, \etc)
were taken from the 1998 Particle Data Group (PDG) tables. 
More precise values for these branching fractions and 
constants are now available from Refs.~\cite{HFAG,PDG}, and,
using these, it is thus possible to, at least partially, update the OPAL
spectral functions. Carrying out this update, and refitting the resulting
modified weighted spectral integrals using the methods
developed and tested in \paper\ is the primary aim of the present article.

We have also investigated the probability distribution of the model parameters that underlies the various
fits to the OPAL data in much more detail, using a Markov-chain Monte Carlo
(McMC)
code in order to map out the {\it a posteriori} distribution.   This is
useful since the fits we perform are non-linear in the parameters, so that
not much is known {\it a priori} about the shape of the probability distribution.
This exploration helps with understanding various potential instabilities in the 
fits (as already detected in \paper), as we will discuss in detail below.

As in \paper, we carry out the analysis using both fixed-order perturbation theory
(FOPT) and contour-improved perturbation theory (CIPT) \cite{CIPT}.\footnote{For recent investigations of these two resummation schemes, see
Refs.~\cite{BJ,SM,CF,DM,AAC}.}
In both cases, we find that the central values for $\a_s$ increase
compared to the values found in \paper, though the two sets of values
are consistent within errors.   The errors themselves stay approximately
the same, which is no surprise, because they are primarily determined by 
the errors on the OPAL spectral data.   

In Sec.~\ref{theory} we briefly review the essentials of the theory needed
to understand the parametrization used in our fits to the OPAL data.
In Sec.~\ref{OPALupdate} we explain in detail how we used recent results from the
Heavy Flavor Averaging Group (HFAG) \cite{HFAG} to update the OPAL spectral functions.  Some details are relegated to an appendix.  Then, in Sec.~\ref{chi2} we discuss what can be learned from the posterior probability distribution obtained with the McMC code.
We present the results of our fits in
Sec.~\ref{fits} and summarize them in Sec.~\ref{summary}.   In Sec.~\ref{future},
we argue that the reduction of errors on the spectral functions expected from the
BaBar or Belle data are likely to be of significant help in reducing the non-perturbative uncertainties.
Section~\ref{conclusion} contains our conclusions.

%\newpage
\section{\label{theory} Theoretical parametrization}
%%####%%
We start with a very brief review of the theory underlying our fits,
referring the reader to \paper\ for more details.   Our fits are based on
FESRs of the form \cite{shankar,BNP}
\begin{equation}
\label{FESR}
I^{(w)}_{V/A}(s_0)\equiv\int_0^{s_0}\frac{ds}{s_0}\;w(s)\;\r^{(1+0)}_{V/A}(s)
=-\frac{1}{2\p i}\oint_{|s|=s_0}\frac{ds}{s_0}\;w(s)\;\P^{(1+0)}_{V/A}(s)\ ,
\end{equation}
where the weight $w(s)$ is a polynomial in $s$, and 
$\P^{(1+0)}_{V/A}(s)$ with $s=q^2=-Q^2$ is defined by
\begin{equation}
\label{corr}
i\int d^4x\,e^{iqx}\,\langle 0|T\left\{J_\m(x)J_\n^\dagger(0)\right\}|0\rangle
=\left(q_\m q_\n-q^2 g_{\m\n}\right)\P^{(1+0)}(s)+q^2 g_{\m\n}\P^{(0)}(s)\ .
\end{equation}
Here $J_\m$ is one of the non-strange $V$ or $A$ currents
$\bu\g_\m d$ or $\bu\g_\m\g_5 d$, and the superscripts $(0)$ and $(1)$
label spin.  

The spectral functions $\r^{(1+0)}_{V/A}$ are taken from OPAL \cite{OPAL},
and the integral on the left-hand side of Eq.~(\ref{FESR}) is then approximated
by a sum over bins, with $s_0\in[s_{min},s_{max}]$, which is our fitting
interval.   These data do not contain the pion pole, which
needs to be added by hand. Other (pseudo-)scalar contributions are 
numerically negligible, being
suppressed by two powers of the light quark masses,\footnote{The second $\d$-function in $\r^{(0)}_A(s)$
comes from the kinematical singularity in Eq.~(\ref{corr}). However, the combination
$\r^{(1+0)}$ is free from kinematical singularities.}
\begin{eqnarray}
\label{spin0}
\r^{(0)}_V(s)&=&O[(m_u-m_d)^2]\ ,\\
\r^{(0)}_A(s)&=&2f_\p^2\left(\d(s-m_\p^2)-\d(s)\right)+O[(m_u+m_d)^2]\ .
\nonumber
\end{eqnarray}
In our fits, we will use the value $f_\p=92.21\pm 0.14$~MeV \cite{PDG}.
The right-hand side of Eq.~(\ref{FESR}) provides the connection to theory,
and is parametrized in terms of the strong coupling $\a_s(m_\t^2)$,
the OPE condensates, and a parametrization of the DV part of $\P^{(1+0)}_{V/A}(s)$.
We write (for both $V$ and $A$)
\begin{equation}
\label{th}
\P^{(1+0)}(s)=\P^{(1+0)}_{ \rm pert}(s)+\P^{(1+0)}_{\rm OPE}(s)+\P^{(1+0)}_{\rm DV}(s)\ ,
\end{equation}
with the subscripts ``pert,'' ``OPE,'' and ``DV'' denoting the perturbative,
OPE (of dimension larger than zero), and DV contributions to $\P^{(1+0)}(s)$.

The perturbative part of the right-hand side of Eq.~(\ref{FESR}) can, by
partial integration, be written in terms of the perturbative Adler function
\begin{eqnarray}
\label{Adler}
D^{(1+0)}_{ \rm pert}(s)&=&-s\,\frac{d\P^{(1+0)}_{ \rm pert}(s)}{ds}\\
&=&\frac{1}{4\p^2}\sum_{n=0}^\infty a_s^n(\m^2)\sum_{k=1}^{n+1}
kc_{nk}\left(\log\frac{-s}{\m^2}\right)^{k-1}\ ,\nonumber
\end{eqnarray}
where $a_s(\m^2)\equiv\a_s(\m^2)/\p$.  Since $D(s)$ is independent
of $\m$, we can choose (for instance) $\m^2=s_0$ in Eq.~(\ref{FESR}), which corresponds
to the FOPT scheme, or $\m^2=-s$, which corresponds to the CIPT
scheme \cite{CIPT}.   We will use values for the coefficients $c_{n1}$
calculated in Ref.~\cite{PT4loop} up to order $n=3$ and in Ref.~\cite{PT} up to order $n=4$; for $c_{51}$ we use the
estimate $c_{51}=283\pm 283$ of Ref.~\cite{BJ}.   The values of $c_{nk}$ for
$k>1$
follow from the $c_{n1}$ using a renormalization-group analysis based
on the fact that the Adler function is independent of $\m$ \cite{MJ}.

The (higher-dimension) OPE contribution can be expressed in terms of
the OPE coefficients $C_{D=2k}$ as
\begin{equation}
\label{OPE}
\P^{(1+0)}_{\rm OPE}(s)=\sum_{k=1}^\infty\frac{C_{2k}(s)}{(-s)^k}\ .
\end{equation}
In our fits we will set $C_2=0$ (it is purely perturbative and suppressed
by two powers of the light quark masses),\footnote{For a alternative view
of the $D=2$ contribution in this context, see Refs.~\cite{NZ,SN}.} and we will treat $C_4$, $C_6$
and $C_8$ as constant, neglecting logarithmic $s$ dependence; we will have no need for the coefficients $C_{D>8}$.
To leading order in $\a_s$, and ignoring tiny isospin-breaking effects and
perturbative light-quark mass contributions, $C_4$ is the same in the $V$ and
$A$ channels; this is not the case for $C_6$ and $C_8$.   
For a more detailed discussion, including references, see \paper.

Finally, the DV contribution to the right-hand side of Eq.~(\ref{FESR}) can be
expressed in terms of the DV part of the spectral function 
\begin{equation}
\label{specDV}
\r^{\rm DV}(s)=\frac{1}{\p}\,\mbox{Im}\,\P^{(1+0)}_{\rm DV}(s)\ ,
\end{equation}
as \cite{CGP}
\begin{equation}
\label{FESRDV}
\cd_w(s_0)=-\frac{1}{2\p i}\oint_{|s|=s_0}\frac{ds}{s_0}\;w(s)\;\P^{(1+0)}_{\rm DV}(s)
=-\int_{s_0}^\infty \frac{ds}{s_0}\,w(s)\,\r^{\rm DV}(s)\ .
\end{equation}
In a slight variation on Ref.~\cite{CGP}, we parametrize $\r^{\rm DV}_{V/A}$ as
\begin{equation}
\label{DVpar}
\r^{\rm DV}_{V/A}(s)=\mbox{exp}\left(-\d_{V/A}-\g_{V/A}s\right)\sin\left(\a_{V/A}+\b_{V/A}s\right)\ .
\end{equation}
This adds four new parameters per channel, in addition to $\a_s$ and the OPE coefficients, to the fits to Eq.~(\ref{FESR}).   
The interval $[s_{min},s_{max}]$ has to be chosen such that the expressions~(\ref{Adler}),
~(\ref{OPE}) and~(\ref{FESRDV}) with~(\ref{DVpar})
provide an accurate representation of the right-hand side of Eq.~(\ref{FESR})
over the whole interval.
The \ansatz~(\ref{DVpar}) was developed in Refs.~\cite{CGP05,CGPmodel}, based on the earlier ideas of Ref.~\cite{russians}.\footnote{The parametrization of DVs is also discussed in Ref.~\cite{MJDV}.}

In Eq.~(\ref{DVpar}), we have traded the parameters $\k_{V/A}$ of \paper\ for the
parameters $\d_{V/A}$; they are related (for both $V$ and $A$) by
\begin{equation}
\label{logkappa}
\k=e^{-\d}\ .
\end{equation}
The reason for making this change is that the fit errors on $\d$ 
are much more symmetric than those on $\k$.  
The (strong) correlations between $\k$ and $\g$ in each channel 
correspond to correlations between
$\d$ and $\g$ which 
are much closer to linear.

In this article, as in \paper, we will employ the weights
\begin{eqnarray}
\label{weights}
\hw_0(x)&=&1\ ,\\
\hw_2(x)&=&1-x^2\ ,\nonumber\\
\hw_3(x)&=&(1-x)^2(1+2x)\ ,\nonumber\\
x&\equiv&s/s_0\ .\nonumber
\end{eqnarray}
The weight $\hw_3$ corresponds
to the (spin-1) kinematic weight that 
appears in the hadronic branching ratio $R_\t$.
Note that 
\begin{equation}
\label{Rt}
R^{(1+0)}_{V+A,ud}(s_0)=12\p^2S_{EW}|V_{ud}|^2I^{(\hw_3)}_{V+A}(s_0)
\end{equation}
is, for $s_0=m_\t^2$, equal to the $(1+0)$ contribution to the ratio of the
non-strange hadronic decay width and the electronic decay width of the
$\t$.  In the following, we will find it convenient to distinguish between
$I^{(w)}_{\rm ex}(s_0)$, denoting the left-hand side, and
$I^{(w)}_{\rm th}(s_0)$, denoting the right-hand side of Eq.~(\ref{FESR}).

If we choose $C_{4}$, $C_{6,V/A}$ and $C_{8,V/A}$ constant, it 
follows that none of these coefficients contribute to $I^{(\hw_0)}_{\rm th}$, only $C_{6,V/A}$ contribute to $I^{(\hw_2)}_{\rm th}$, and both $C_{6,V/A}$ and $C_{8,V/A}$ contribute to $I^{(\hw_3)}_{\rm th}$.\footnote{We have 
checked the influence of higher-order $\a_s$ corrections  to the $D=4$ contributions in the OPE.   Numerically, the differences
are tiny, and can safely be neglected.  For more discussion, see~\paper.}   Since we will not
use weights of degree larger than 3, there is no need to consider the
OPE coefficients $C_D$ with $D>8$.

Weights $w(x)$ which are functions of the dimensionless variable 
$x=s/s_0$ are chosen in order to facilitate the separation of OPE 
contributions to $I_{th}^{(w)}(s_0)$ having different $D=2k$ which, with 
this choice, scale as $1/s_0^k$. While non-perturbative contributions
are small at the scales of typical $\tau$-decay analyses, at the
level of precision claimed in recent $\alpha_s$ determinations
they are definitely {\it not} negligible.   For example, almost the entire difference between
the results of Refs.~\cite{Davieretal08} and \cite{MY} are due to differences in the fitted non-perturbative 
contributions. As discussed in detail in Ref.~\cite{MY} and \paper,
taking advantage of the $s_0$-dependence of the moments $I_{ex}^{(w)}(s_0)$
is crucial for properly constraining such higher-$D$ contributions.
For further discussion of the selection of the particular set of
weights chosen above
we refer the reader to \paper.  

%\newpage
\section{\label{OPALupdate} The OPAL data update}
%%####%%
The 1998 OPAL inclusive $\rho_V$ and $\rho_A$ distributions were 
constructed as sums over exclusive mode distributions.
In this process, the distributions of the three main hadronic
modes in each channel ($\pi^-\pi^0$, $\pi^-3\pi^0$ and 
$\pi^-\pi^+\pi^-\pi^0$ for the $V$ channel and $\pi^-2\pi^0$,
$\pi^-\pi^+\pi^-$ and $\pi^-\pi^+\pi^-2\pi^0$ for the $A$ channel)
were explicitly measured, while the small residual contributions 
associated with other modes were typically Monte Carlo generated using 
TAUOLA 2.4 \cite{OPAL}. The normalizations of the exclusive modes 
(residual or not) were, however, not measured by OPAL, but rather fixed by the 
1998 PDG values for the exclusive-mode branching fractions. 
Significant improvements to these branching fractions
have been made since 1998. 

Since the distributions for the main exclusive modes noted above are
publicly available, it is possible to update the dominant
contributions to the inclusive $V$ and $A$ distributions
by simply rescaling these contributions with the ratio of the new and old branching
fractions for these exclusive modes. Unfortunately,
this is not the case for the residual mode contributions since the 
individual Monte-Carlo-generated residual exclusive-mode distributions
are not publicly available. The distribution for the sum of
residual modes in each channel is, however, reconstructable from
the publicly accessible inclusive and exclusive mode distributions. This
distribution may then be updated in an averaged sense by computing
the new and old versions of the sum of residual-mode branching
fractions and rescaling the old combined residual-mode distribution by the
ratio of these results. Since different exclusive modes have
different $s$-dependent distributions, this average 
updating of the residual distributions is not perfect.
Fortunately, however, the residual modes do not play a major role in
the spectral functions in the kinematically accessible region
(accounting, for example, for only $2.6\%$ of the inclusive
branching fraction in the $V$ channel and only $1.7\%$ in
the $A$ channel). The average rescaling required for the combined 
$V$-channel residual branching fraction turns out to be 
small (reducing the OPAL combined residual-mode branching fraction 
sum by only $1.7\%$).
In contrast, the HFAG version of the $A$-channel combined
residual branching fraction is $1.394$ times the corresponding
OPAL value, making the average residual-distribution rescaling procedure
much safer for the $V$ channel than it is for the $A$ channel.

We perform the updates of both exclusive mode distributions and the 
combined residual mode distributions using branching fractions from a 
recent unitarity-constrained HFAG fit.\footnote{The updated OPAL data are available on request.}
 The particular fit we employ is 
that incorporating Standard Model expectations based on $\pi_{\mu 2}$ 
and $K_{\mu 2}$ data for $B[\tau\rightarrow \pi\nu_\tau ]$ and 
$B[\tau\rightarrow K\nu_\tau ]$ in addition to the results for these
branching fractions measured directly in $\tau$ decays~\cite{HFAG}.\footnote{
We refer to
http://www.slac.stanford.edu/xorg/hfag/tau/hfag-data/tau/2009/TauFit\_Mar2011
/BB\_PiKUniv/ConstrainedFit.pdf for details.} 

It is important to note that the conventions for quoting
the various exclusive branching fractions are not identical for
OPAL and HFAG. HFAG quotes $\omega\pi^-$, $\omega\pi^-\pi^0$ and
$\eta\pi^-\pi^0$ branching fractions corresponding to all
$\omega$ and $\eta$ decay modes, and excludes $\omega$ and $\eta$
substate contributions in quoting branching fractions for all other
modes. In contrast, for OPAL, (i) the quoted $\pi^-\pi^+\pi^-$,
$\pi^-\pi^+\pi^-\pi^0$ and $\pi^-\pi^+\pi^- 2\pi^0$ branching fractions 
include, respectively, $\omega\pi^-$, $\omega\pi^-$ and $\omega\pi^-\pi^0$, 
and $\omega\pi^-\pi^0$ and $\eta\pi^-\pi^0$ components, and 
(ii) the $\omega\pi^-$ and $\omega\pi^-\pi^0$ 
branching fractions are quoted excluding $\omega\rightarrow 3\pi$ 
contributions. With these 
conventions, the tabulated OPAL exclusive branching fractions and 
distributions include small ```wrong-current contaminations'' 
associated with isospin-breaking $\omega\rightarrow\pi^+\pi^-$ and 
$\eta\rightarrow \pi^+\pi^-\pi^0$ decays. Explicitly,
$\omega\rightarrow\pi^+\pi^-$ decays cause the $V$-current-induced 
$\omega\pi^-$ mode to populate the nominally $A$-current $\pi^-\pi^+\pi^0$
distribution and the $A$-current-induced $\omega\pi^-\pi^0$ mode
to populate the nominally $V$-current $\pi^-\pi^+\pi^-\pi^0$ 
distribution, while $\eta\rightarrow\pi^+\pi^-\pi^0$ decays cause
the $V$-current-induced $\eta\pi^-\pi^0$ mode to populate the nominally $A$-current $\pi^-\pi^+\pi^- 2\pi^0$ distribution. In forming the inclusive
$V$ and $A$ spectra, OPAL corrects for this contamination by including an
appropriate negatively weighted version of the relevant
non-$\omega\rightarrow 3\pi$ and non-$\eta\rightarrow 3\pi$ 
$\omega\pi^-$, $\omega\pi^-\pi^0$ and $\eta\pi^-\pi^0$
distributions in the wrong-current inclusive distribution sum. The negative
weights employed by OPAL were determined using the 1998 PDG values
for the branching fractions of the relevant $\eta$ and $\omega$ decay
modes. 

In order to perform the rescaling of the OPAL exclusive-mode distributions, 
the relevant updated wrong-current contaminations must be added to the HFAG 
exclusive branching fractions. The HFAG-updated 
$\omega\pi^-$ (excluding $\omega\rightarrow 3\pi$), 
$\omega\pi^-\pi^0$ (excluding $\omega\rightarrow 3\pi$),  
and $\eta\pi^-\pi^0$ (excluding $\eta\rightarrow 3\pi$)
branching fractions, and updated negative-weight,
wrong-current contamination corrections must, analogously, be incorporated
in the updated version of the combined residual mode
branching fractions in both channels. These updates are performed
using the HFAG exclusive branching fractions, together with 2010 PDG results
for the relevant $\eta$ and $\omega$ branching fractions. Numerical
details may be found in the Appendix.

OPAL has also tabulated the covariance matrices for the three
main exclusive modes in each channel, as well as the $VV$, $VA$ and
$AA$ covariances for the inclusive $V$ and $A$ distributions.
The absence of information on the covariances among the different
exclusive mode distributions limits our ability to update the
inclusive $VV$, $VA$ and $AA$ covariances. Updates for improvements 
in factors such as $B_e$ and $V_{ud}$ which enter when converting
the differential branching fraction distributions,
$dB_{V/A}(s)/ds$, to the corresponding spectral functions, can,
however, be performed. Details on carrying out this procedure
may also be found in the Appendix.

%\newpage
\section{\label{chi2} The posterior probability distribution}
%%####%%
The fit functions used in the sum rules~(\ref{FESR}) are non-linear in 
$\a_s(m^2_\t)$ and the DV parameters.   It is therefore not obvious what the
posterior probability  distribution of the model parameters looks like, even if we assume the
data errors to follow a (multivariate) gaussian distribution.

In order to study this distribution, we have used an McMC code, Hrothgar
\cite{Andi},
in order to generate the conditional probability distribution, which we take to be proportional to $\mbox{exp}[-\c^2({\vec p})/2]$,
given the data, where $\vec p$ represents the array of fit parameters.
With the data fixed, these parameters are varied stochastically, and a 
Metropolis-Hastings accept-reject step is used to generate a statistical picture of the
probability distribution.  In this section, we will describe our findings in more
detail for the case of a fit to the FESR with weight $\hw_0$, first in the $V$ channel.  

The McMC code generates points in the 5-dimensional space spanned by the
five parameters $\a_s(m^2_\t)$, $\d_V$, $\g_V$, $\a_V$ and $\b_V$, and also
computes the value of $\c^2$ at each of the generated points.    These points
are distributed following $\mbox{exp}[-\c^2({\vec p})/2]$, with $\c^2({\vec p})$ evaluated on the (updated)
OPAL data (including the full covariance matrix) and the values of the parameters $\vec p$
at these
points.

%%%%%%%%%%%%%%%%%%%
\begin{figure}
%\vspace*{4ex}
\begin{center}
\includegraphics*[width=12cm]{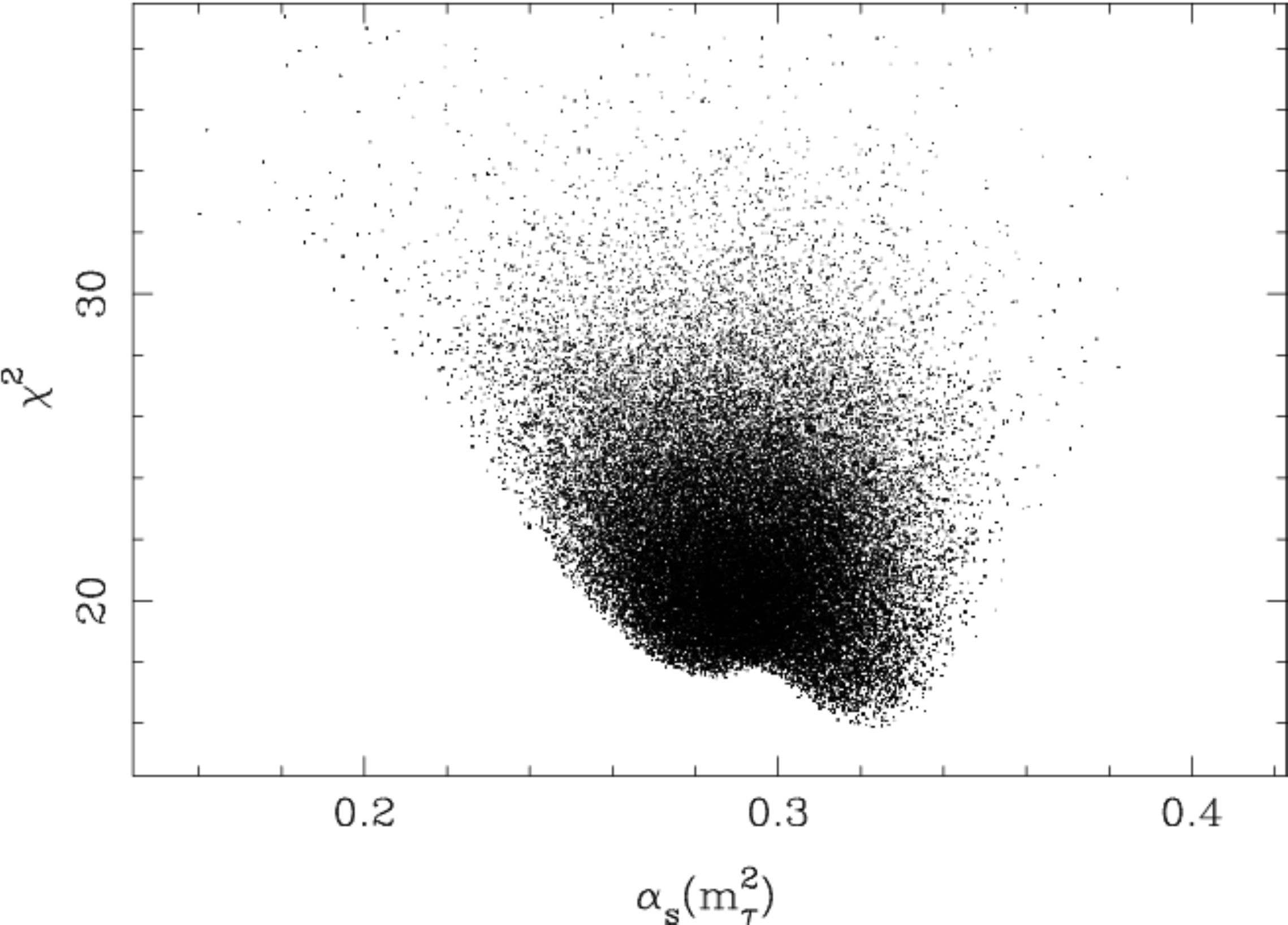}
\end{center}
\begin{quotation}
\floatcaption{as-chi2}%
{{\it $\a_s(m^2_\t)$ versus  $\c^2$; $V$ channel, with $s_{min}=1.5$~{\rm GeV}$^2$ (200,000 points).}}
\end{quotation}
\vspace*{-4ex}
\end{figure}
%%%%%%%%%%%%%%%%%%%

The probability distribution thus obtained can be projected onto two-dimensional planes.   In Fig.~\ref{as-chi2}, we show $\c^2$ as a function of
$\a_s(m^2_\t)$, choosing $s_{min}=1.5$~GeV$^2$, using FOPT for the
perturbative part.\footnote{The distribution for CIPT looks essentially the same, except that the projections shown in Figs.~\ref{as-chi2} 
and~\ref{Vw0proj}, left panel, are shifted to the right by an amount 
$\sim 0.02$.}   Since for each
$\a_s(m^2_\t)$ points with many different values for the other four parameters are generated stochastically, the distribution appears as the cloud shown in the figure.\footnote{If a new point is rejected by the accept-reject step, the old
point is retained.   Therefore each point in the plot may represent multiple points in the ensemble.}   

%%%%%%%%%%%%%%%%%%%
\begin{figure}[t]
\centering
\includegraphics[width=2.9in]{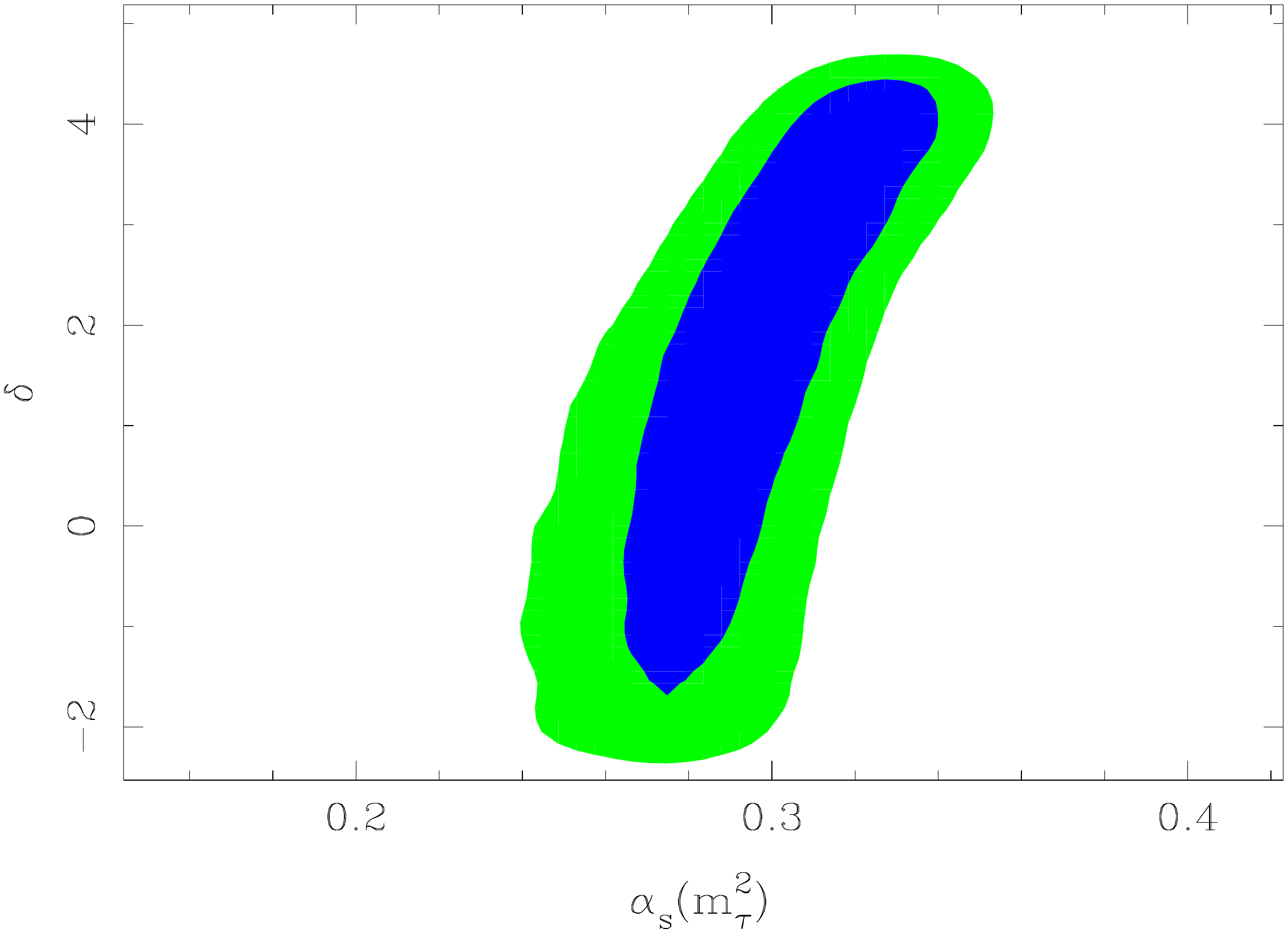}
\hspace{.1cm}
\includegraphics[width=2.9in]{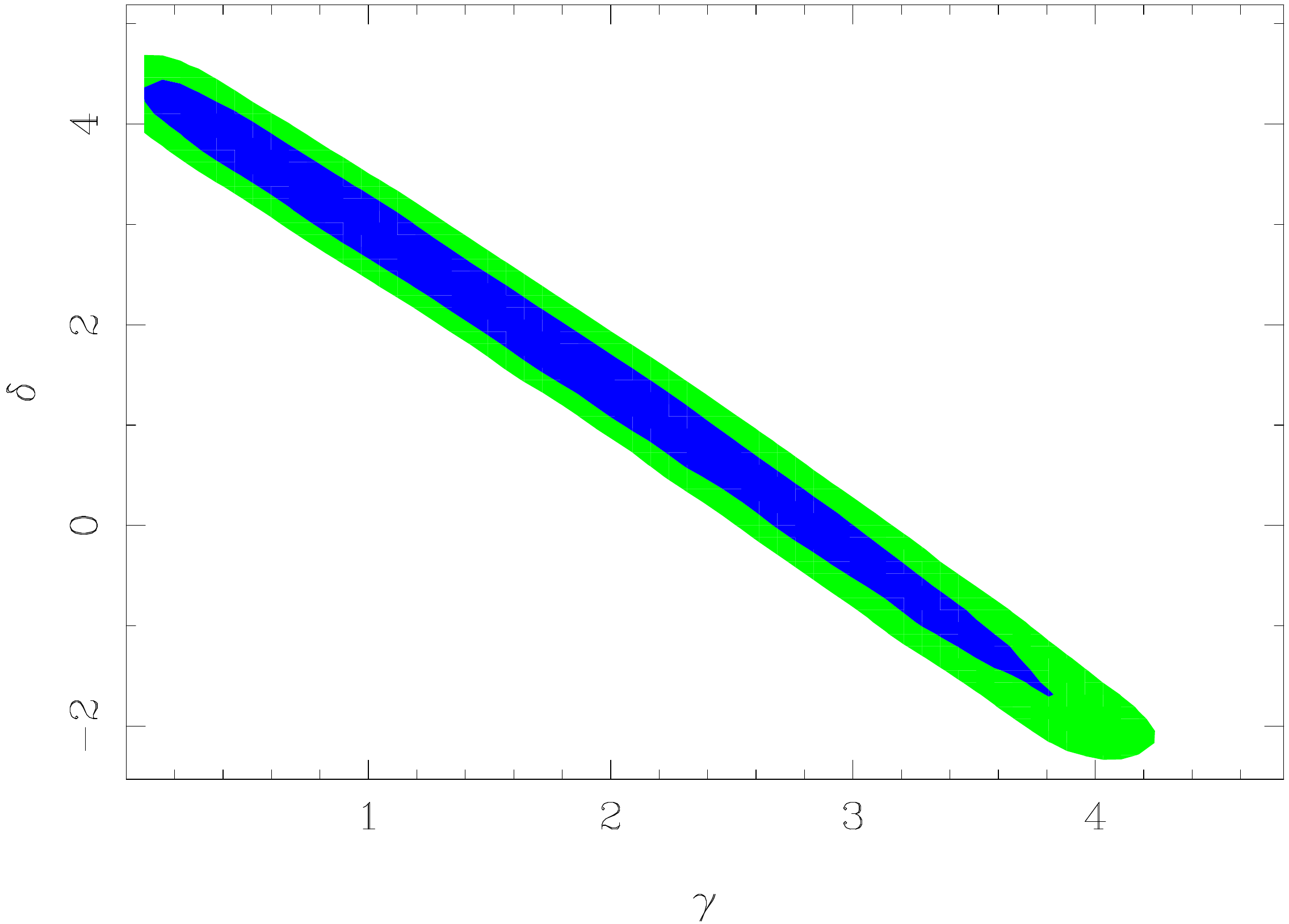}
\floatcaption{Vw0proj}{\it 
Two-dimensional contour plots showing $\a_s(m^2_\t)$ versus $\d_V$ and
$\g_V$ versus $\d_V$.  Left panel: projection onto $\a_s(m^2_\t)-\d_V$ plane.   Right panel: projection onto $\g_V-\d_V$ plane.  $V$ channel, $s_{min}=1.5\ {\rm GeV}^2$.  Blue (darker) areas and green (lighter) areas contain
68\%, respectively, 95\% of the distribution.}
\vspace*{2ex}
\end{figure}
%%%%%%%%%%%%%%%%%%%

Figure~\ref{as-chi2} shows a bi-modal distribution, with one local minimum near 
$\a_s(m^2_\t)=0.28$, and a global minimum near $\a_s(m^2_\t)=0.32$,
with a difference in the locally minimal values of $\c^2$ equal to about $1.6$.   As a 
consequence, a standard $\c^2$ minimization, which estimates the 
parameter covariance matrix from the hessian at the (global) minimum,%
\footnote{Or from the minimum value of $\c^2$ plus one.} will miss the other
local minimum entirely.

The origin of the problem appears to be the fact that $\d_V$ is not well constrained by the data.   This can be seen in Fig.~\ref{Vw0proj}, which 
shows the projections onto the $\a_s(m^2_\t)-\d$ and
$\g-\d$ planes, in this case as contour plots showing the regions containing
68\% (blue) and 95\% (green) of the distribution.  The right panel 
shows a very strong correlation between the two parameters, $\d_V$ and $\g_V$,
which together control the ``strength'' of the DV part of the spectral functions
in the low-$s$ part of our fitting windows,
\seef\ Eq.~(\ref{DVpar}).  Clearly, external input is required to narrow down which part of the distribution is most likely to correspond to physics.
This will be discussed  in Sec.~\ref{fits}.

%%%%%%%%%%%%%%%%%%%
\begin{figure}[t]
%\vspace*{4ex}
\begin{center}
\includegraphics*[width=7cm]{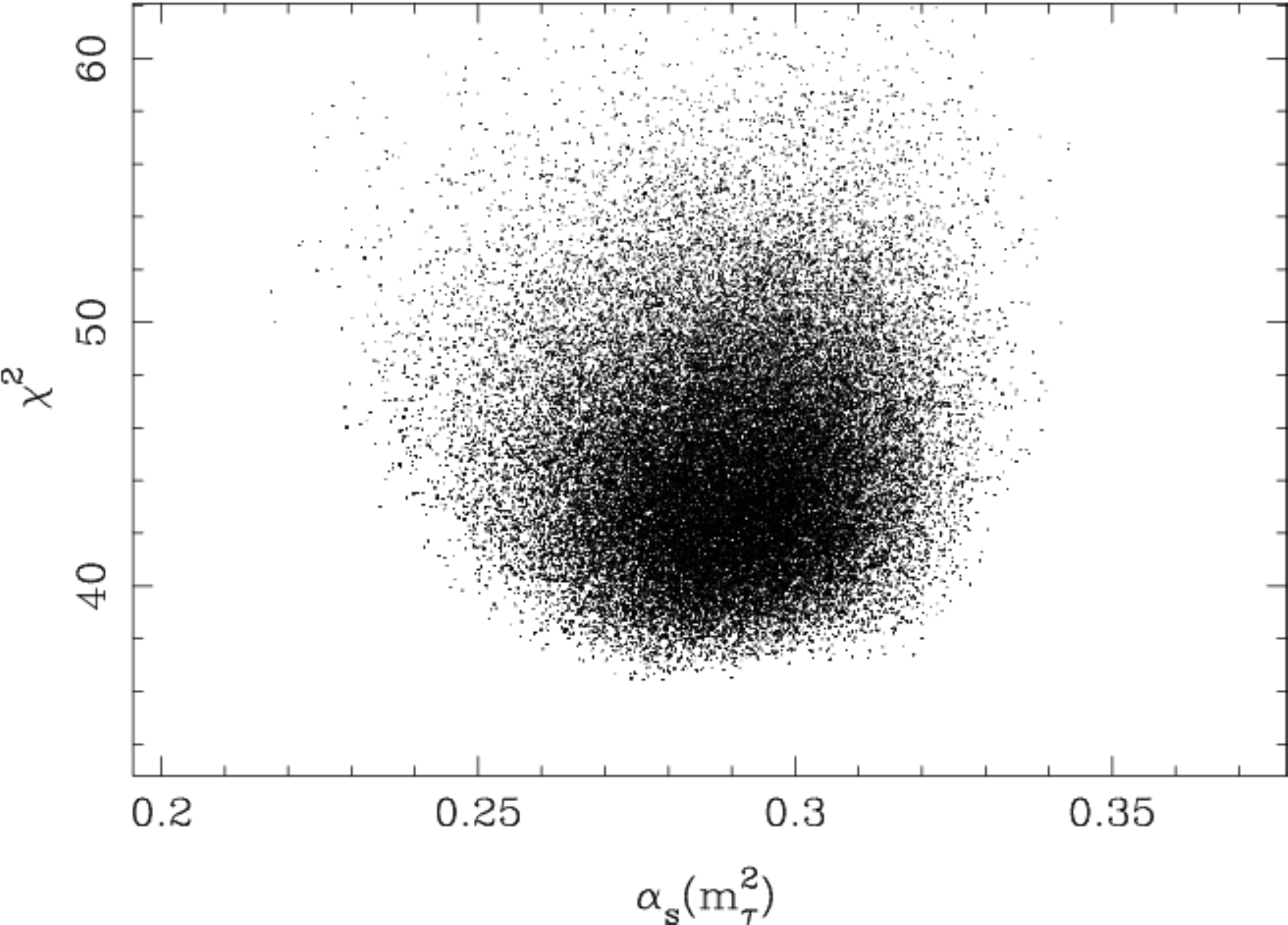}
\hspace{.1cm}
\includegraphics*[width=7cm]{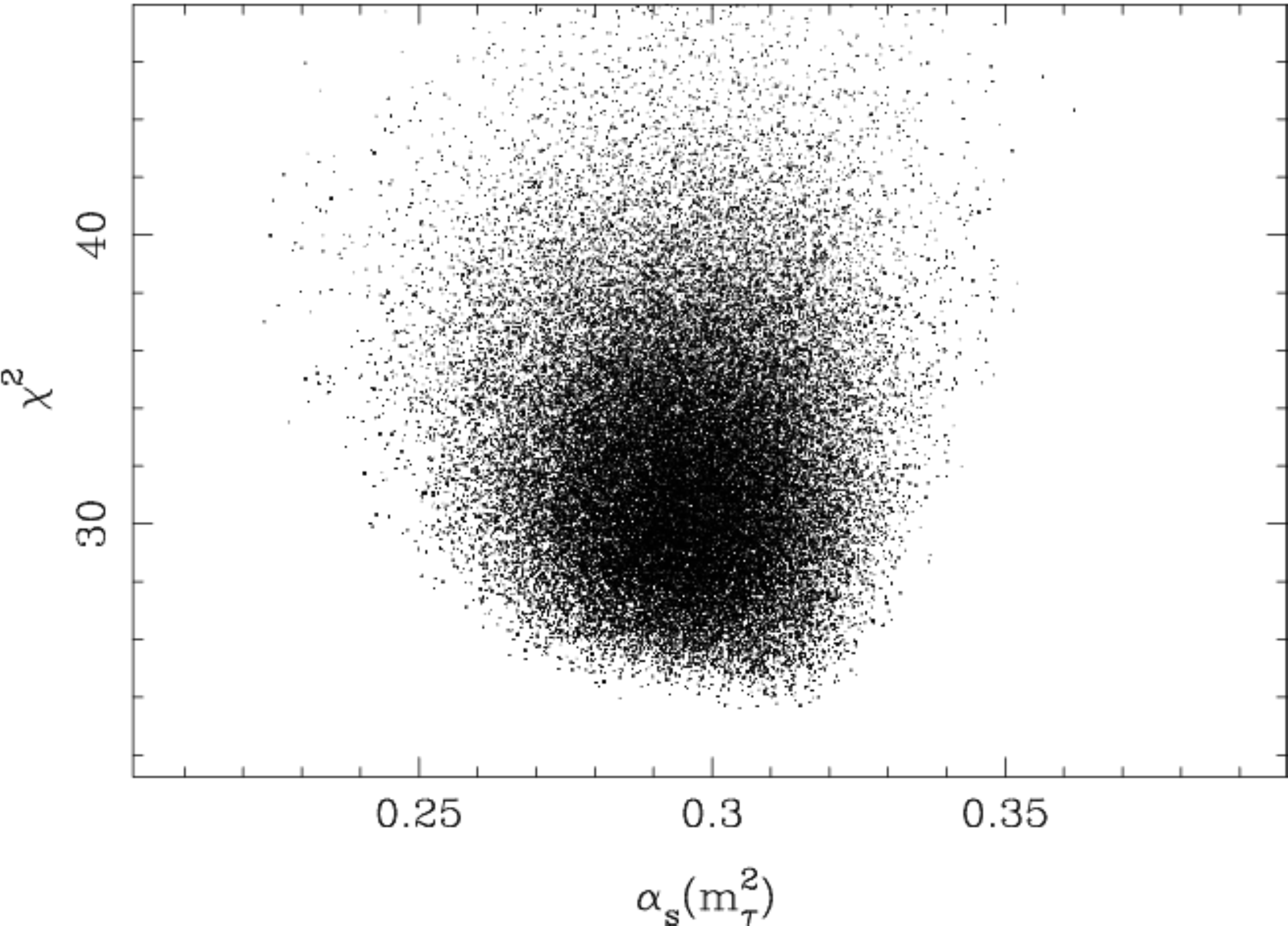}
\includegraphics*[width=7cm]{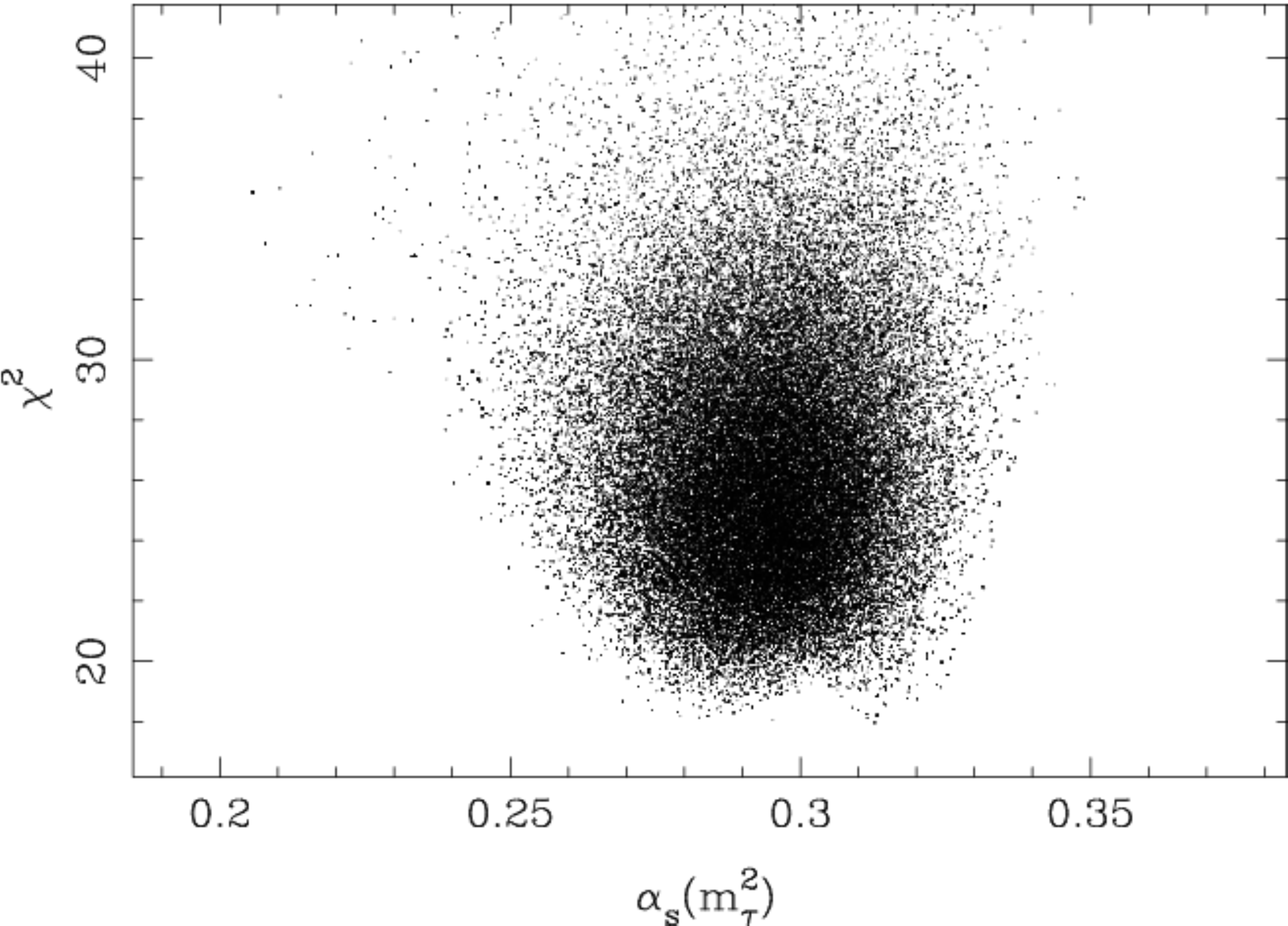}
\hspace{.1cm}
\includegraphics*[width=7cm]{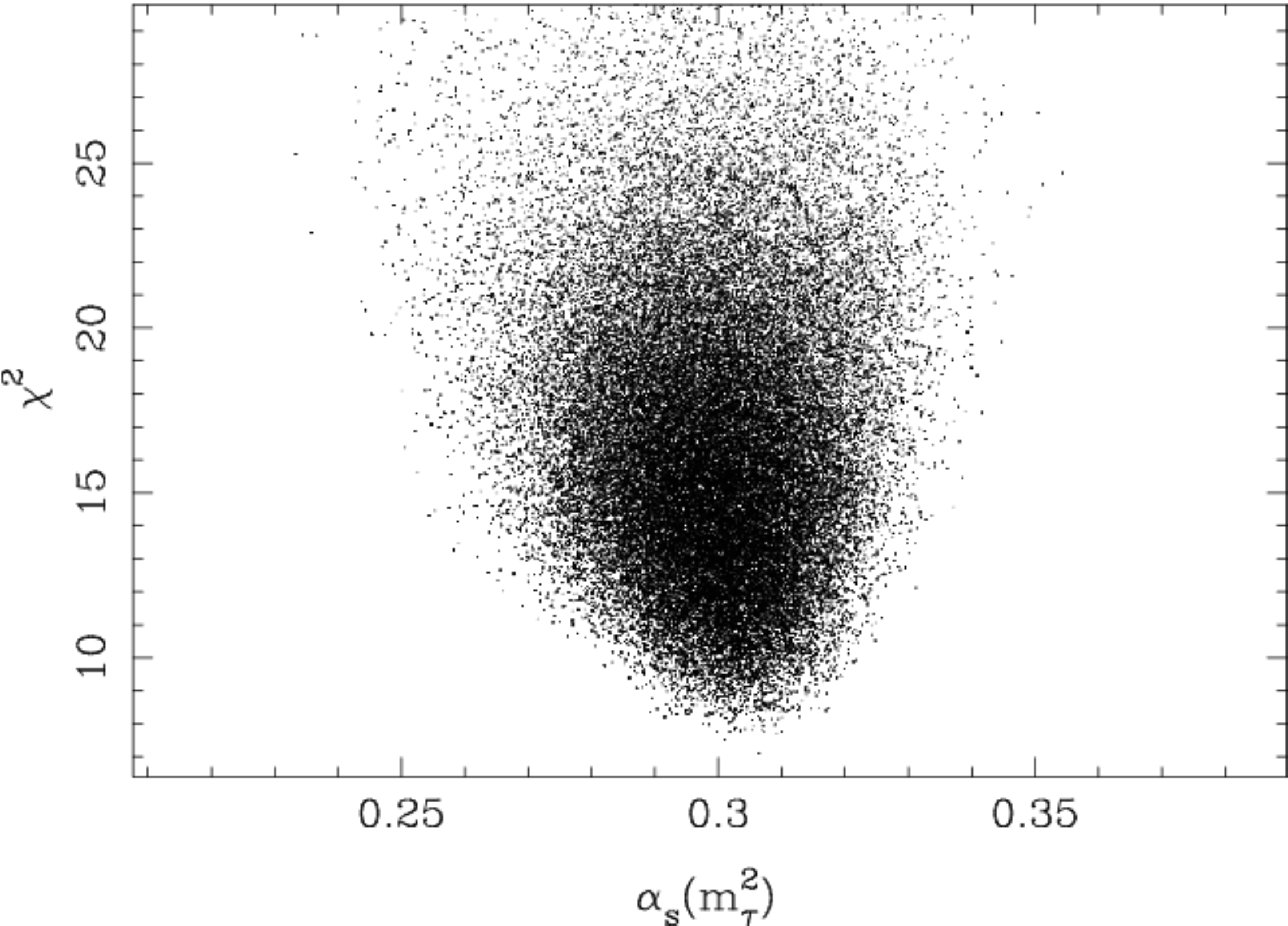}
\end{center}
\begin{quotation}
\floatcaption{as-chi2VA}%
{{\it $\c^2$ versus $\a_s(m^2_\t)$, thinning out the integrated data by factors 1 (upper left), 2 (upper right), 3 (lower left) and 4 (lower right); $V$ and $A$ channel combined, with $s_{min}=1.5$~{\rm GeV}$^2$ (200,000 points).}}
\end{quotation}
\vspace*{-4ex}
\end{figure}
%%%%%%%%%%%%%%%%%%%

Figure~\ref{as-chi2VA}  shows analogous results for
a $V$ and $A$ channel combined ($V\&A$) fit, again for $s_{min}=1.5$~GeV$^2$, again using FOPT.  In this case, there are 9 fit parameters, and the figure shows a projection of the 10-dimensional space spanned by the 9 parameters and
$\c^2$.
The four panels show a fit of the $V\&A$  FESR with moment $\hw_0$,
similar to the $V$ channel fit shown in Fig.~\ref{as-chi2}.  In the upper left panel we used the full set of $s_0$ values corresponding to the right end-points of all bins starting from $s_{min}=1.504$~GeV$^2$, whereas the other three panels
show fits in which the $s_0$ values employed have been thinned out by a ``thinning factor'' $n$,
chosen equal to 2, 3 and 4, respectively, in the upper right, lower left and
lower right panels.\footnote{If the thinning factor is equal to $n$, we use every $n$-th value of $I^{(\hw_0)}_{\rm ex}(s_0)$ in the fit, see also \paper.
We emphasize that all data are used: only integrated data are thinned out.}
Contour plots for the combinations $\a_s(m^2_\t)-\delta_V$ and $\g_V-\d_V$ look very similar to those shown for the $V$ case in Fig.~\ref{Vw0proj}.

Again, as in Fig.~\ref{as-chi2}, there appear to be two local minima, one
centered around $\a_s(m^2_\t)=0.315$, and one centered around
$\a_s(m^2_\t)=0.28$.   However, in this case the two minima are much closer
to being degenerate than in the $V$-channel fit.
For $n=4$ it is difficult to discern two separate minima;
the two minima appear to merge.

Similar behavior as a function of the thinning factor is observed in fits to only the $V$ channel as well, but the two 
minima are always more clearly separated, as in Fig.~\ref{as-chi2}.  This may explain why the $V\&A$ combined fits in \paper\ were found to be less stable than  $V$ channel fits.   Figure~\ref{as-chi2VA} may also explain why fits with $n=3$ led to more stable
results in \paper, since for $n=3$ the two minima appear to be somewhat 
more clearly separated than for other values of $n$.\footnote{We have checked that the behavior of the posterior probability distributions with the non-updated data we used in \paper\ is very similar to what we find with the updated data.}

One should also bear in mind, in assessing the relative
reliability of the results of the $V$ channel and combined
$V\& A$ channel fits, that the much larger
average rescaling of the sum-of-residual-modes branching fraction
in the $A$ channel makes the updating of the spectral function
much less reliable in the $A$ channel than it is in the $V$ channel.

For both the $V$-only and the $V\&A$ fits, we have also studied the behavior of the $\c^2$ distribution as a function of $s_{min}$.   We find that by lowering $s_{min}$, the minimum at the lower value of $\a_s(m^2_\t)$ ``moves up,'' \ie, the value of $\c^2$ at that local minimum increases relative to that
at the other minimum.  

%\newpage
\section{\label{fits} Fits}
%%####%%
In this section, we present the 
results of fits to a range of different moments, obtained
by minimizing various different ``fit qualities,'' (positive-definite 
quadratic forms in the differences between theory and data).
In Sec.~\ref{w=1} we discuss fits to the FESR with moment
$I^{(\hw_0)}$, whereas in Sec.~\ref{moments} we will consider simultaneous fits 
to FESRs with the moments $I^{(\hw_{0,2,3})}$, using the weights of 
Eq.~(\ref{weights}).  

In the first case, we choose the fit quality to be the standard $\c^2$, 
already discussed in the previous section, constructed with the complete 
(updated) covariance matrix. 
In the second case, in which we combine more than one moment, 
it turns out, as discussed in \paper, that the correlations are 
too strong to allow for a fit based on the standard $\c^2$ 
function.\footnote{When more than one weight is employed,
the correlation matrix for the full set of weighted spectral integrals,
labelled by the weights and $s_0$ values employed, acquires a number of machine-precision
zero eigenvalues.} We therefore employ a somewhat simpler fit quality $Q^2$.
Working with a set of values of $s_0$, $\{s_0^k\}$ in some fitting window,
we define
\begin{equation}
\label{Q2}
Q^2=\sum_w\sum_{s_0^i,\, s_0^j}
\left(I_{ex}^{(w)}(s_0^i)-I_{th}^{(w)}(s_0^i;{\vec p})\right)
\left(C^{(w)}\right)^{-1}_{ij}
\left(I_{ex}^{(w)}(s_0^j)-I_{th}^{(w)}(s_0^j;{\vec p})\right)\ ,
\end{equation}
with $C_w$ the covariance matrix for moments with fixed weight $w$
and $s_0$ running over the chosen fit window.\footnote{The fit quality
$Q^2$ corresponds to $Q^2_{block}$ defined in \paper.}   
The fit quality $Q^2$ is thus similar to a standard $\c^2$, but with 
cross-correlations between different moments omitted. Treating
$Q^2$ as if it were the standard $\c^2$ would thus lead to incorrect
errors and covariances for the fit parameters. To take the 
cross-correlations properly into account, errors and covariances
for $Q^2$-based fits are obtained using the linear fluctuation analysis
described in the Appendix of \paper.

In view of (i) the results of Sec.~\ref{chi2} and
(ii) the fact that the updating scheme is much more reliable for
the $V$ channel than for the $A$ channel OPAL data, we 
will use the $V$-channel fits of Sec.~\ref{w=1}
to determine our central value for $\a_s(m^2_\t)$. 
The remaining fits are used only to investigate whether our 
fit function, which parameterizes DVs using Eq.~(\ref{DVpar}), provides a 
good description of the data for the
moments $\hw_2$ and $\hw_3$ as well. The issue of the choice of weights is discussed in more detail in \paper.

As we have seen in Sec.~\ref{chi2}, the posterior probability distribution generally has a rather complicated structure, showing almost always two fairly close but different minima.
We thus need to address the question which minimum is more likely to correspond to a physical solution.   The situation is more complicated in the case of $V\&A$ fits, for which the two minima are essentially degenerate.   
We will argue that the minimum corresponding to the larger value of $\a_s(m^2_\t)$ is more likely to correspond to the correct physics.  

First, there is evidence for this choice from the fits themselves.   We note that,
 in the $V$-only case, the minimum of $\c^2$ for the larger value 
 of $\a_s(m_\t^2)$ is
always the lower one by a significant amount, \seef\ Fig.~\ref{as-chi2}.   This is confirmed by fits with lower values of $s_{min}$, for which this separation becomes more pronounced.

We may also refer to the model study of Ref.~\cite{CGPmodel}, which led to 
the form of the \ansatz~(\ref{DVpar}) used to parameterize DVs.  
It was shown there that the model underlying this \ansatz\ leads 
naturally to the following values for the parameters:
\begin{equation}
\label{parameters}
    \delta\sim -\log\left(\frac{F^2}{\Lambda^2}\right)\sim 4 \quad \mathrm{and} \quad \gamma \sim \frac{1}{N_c}\frac{1}{\Lambda^2}\sim 0.3\ \mathrm{GeV}^{-2}\  ,
\end{equation}
where $F\sim 0.1$~GeV is a typical value for a resonance decay constant, and $\Lambda\sim 1$~GeV is a typical QCD scale.  
Figure~\ref{Vw0proj} shows that for such values the global $\c^2$ minimum, which occurs at the larger value of $\d_V$, and thus at the larger value of 
$\a_s(m^2_\t)$, is preferred.   (We will see below that the estimates of
Eq.~(\ref{parameters}) are less well satisfied for the $A$ channel.)
Henceforth, we will refer to the minima at larger values of $\d_V$ as 
``physical'' minima, and to those at smaller values of $\d_V$ as ``unphysical.''

\begin{table}[t]
\begin{center}
\begin{tabular}{|c|c|c|c|c|c|c|c|}
\hline
$s_{min}$ & dof &$\c^2$/dof & $\alpha_s$ & $\d_V$ & $\g_V$ & $\a_V$ & $\b_V$ \\
\hline
1.3 & 53 & 0.41 & 0.338(18) & 3.91(62) & 0.27(43) & 0.53(54) & 2.89(29) \\
1.4 & 50 & 0.33 & 0.326(16) & 4.11(63) & 0.16(41) & -0.29(68) & 3.29(35) \\
1.5 & 47 & 0.34 & 0.323(16) & 4.21(62) & 0.12(40) & -0.48(79) & 3.38(40) \\
1.6 & 44 & 0.35 & 0.325(18) & 4.04(86) & 0.20(53) & -0.37(87) & 3.33(44) \\
1.7 & 41 & 0.35 & 0.323(19) & 4.37(99) & 0.05(53) & -0.48(91) & 3.37(44) \\
\hline
\hline
1.3 & 53 & 0.43 & 0.360(32) & 3.47(64) & 0.53(47) & 0.57(58) & 2.83(32) \\
1.4 & 50 & 0.34 & 0.349(25) & 3.84(65) & 0.30(44) & -0.31(67) & 3.28(35) \\
1.5 & 47 & 0.34 & 0.345(24) & 3.99(64) & 0.22(42) & -0.54(77) & 3.39(40) \\
1.6 & 44 & 0.36 & 0.347(26) & 3.82(90) & 0.32(56) & -0.45(86) & 3.35(44) \\
1.7 & 41 & 0.37 & 0.344(25) & 4.2(1.1) & 0.12(57) & -0.57(90) & 3.40(44) \\
\hline
\end{tabular}
\end{center}
\begin{quotation}
\floatcaption{w0}{{\it Standard $\c^2$ fits to
Eq.~(\ref{FESR}) with $w(s)=1$, V channel.
FOPT results are shown above the double horizontal line, CIPT results below.
Errors are standard $\c^2$ errors;
$\g_V$ and $\b_V$ in {\rm GeV}$^{-2}$. }}
\end{quotation}
\vspace*{-4ex}
\end{table}%

%\newpage
\begin{boldmath}
\subsection{\label{w=1} Fits with $\hw_0$}
\end{boldmath}
%%####%%
In Table~\ref{w0} we show $V$ channel fits of $I^{(\hw_0)}_{\rm th}(s_0)$
to $I^{(\hw_0)}_{\rm ex}(s_0)$ (\seef\ Eq.~(\ref{FESR})), for $s_0\in[s_{min},s_{max}]$, with $s_{max}=3.136$~GeV$^2$ and varying
$s_{min}$.\footnote{This value of $s_{max}$ corresponds to the highest bin available in the OPAL data; the bin width is $0.032$~GeV$^2$.
In the axial channel the highest bin available corresponds to $s_{max}=3.104$~GeV$^2$.}   In all the fits contained in this table, we have used initial parameter estimates which roughly correspond to the physical minima, \ie, the minima corresponding to larger values of $\d_V$ found with the McMC code.
There is excellent stability
for $s_{min}$ ranging from 1.4 to 1.7~GeV$^2$.   In a slight deviation from \paper, we will use the average of the fits with
$s_{min}=1.4$, $1.5$ and $1.6$~GeV$^2$ to determine $\a_s(m^2_\t)$.   Since the fit is non-linear,
one expects the fit errors to be asymmetric.  For instance, for the FOPT fit
with $s_{min}=1.5$~GeV$^2$ we find
\begin{eqnarray}
\label{asymm}
\a_s(m_\t^2)&=&0.323^{\,+0.016}_{\,-0.018}\ ,\\
\d_V&=&4.21^{\,+0.53}_{\,-0.88}\ ,\nonumber\\
\g_V&=&0.12^{\,+0.57}_{\,-0.33}\ \mbox{GeV}^{-2}\ ,\nonumber\\
\a_V&=&-0.48^{\,+0.75}_{\,-0.81}\ ,\nonumber\\
\b_V&=&3.38^{\,+0.42}_{\,-0.38}\ \mbox{GeV}^{-2}\ .\nonumber
\end{eqnarray}
We note that the error on $\a_s(m_\t^2)$ is nearly symmetric, and that the
error on $\d_V$ is much closer to symmetric than the error on 
$\k_V={\rm exp}(-\d_V)$ in \paper. A typical parameter correlation matrix, that
for the FOPT fit with $s_{min}=1.5$~GeV$^2$, is shown in 
Table~\ref{w0parcorr}. Results for other values of $s_{min}$, 
or for CIPT fits, show the same pattern.  
\vspace*{1ex}
\begin{table}[h]
\begin{center}
\begin{tabular}{|c|ccccc|}
\hline
 & $\a_s$ & $\d_V$ & $\g_V$ & $\a_V$ & $\b_V$ \\
\hline
$\a_s$ & 1 & 0.68 & -0.67 & 0.74 & -0.68 \\
$\d_V$ & 0.68 & 1 & -0.99 & 0.47 & -0.44 \\
$\g_V$ & -0.67 & -0.99 & 1 & -0.49 & 0.45 \\
$\a_V$ & 0.74 & 0.47 & -0.49 & 1 & -0.98 \\
$\b_V$ & -0.68 & -0.44 & 0.45 & -0.98 & 1 \\
\hline
\end{tabular}
\end{center}
\begin{quotation}
\floatcaption{w0parcorr}{{\it Parameter correlation matrix for the FOPT fit
with $s_{min}=1.5$~{\rm GeV}$^2$ shown in
Table~\ref{w0}.}}
\end{quotation}
\vspace*{-4ex}
\end{table}%

Recalling our choice to obtain a central
value by averaging results for $s_{min}=1.4$, $1.5$ and 
$1.6$~GeV$^2$, we obtain from these fits for $\a_s$ at the $\t$ mass
the results
\begin{eqnarray}
\label{asw0}
\a_s(m_\t^2)&=&0.325\pm 0.016\pm 0.002\pm 0.007\qquad\mbox{(FOPT)}\ ,\\
\a_s(m_\t^2)&=&0.347\pm 0.024\pm 0.002\pm 0.005\qquad\mbox{(CIPT)}\ .\nonumber
\end{eqnarray} 
The first error is the $s_{min}=1.5$~GeV$^2$ fit error shown in Table~\ref{w0},
the second the variation of the central values over 
the $s_{min}= 1.4\rightarrow 1.6$~GeV$^2$ averaging window, and the 
third the result of the $\pm 283$ variation of $c_{51}$
about its central value $c_{51}=283$.

\begin{figure}[t]
\centering
\includegraphics[width=2.9in]{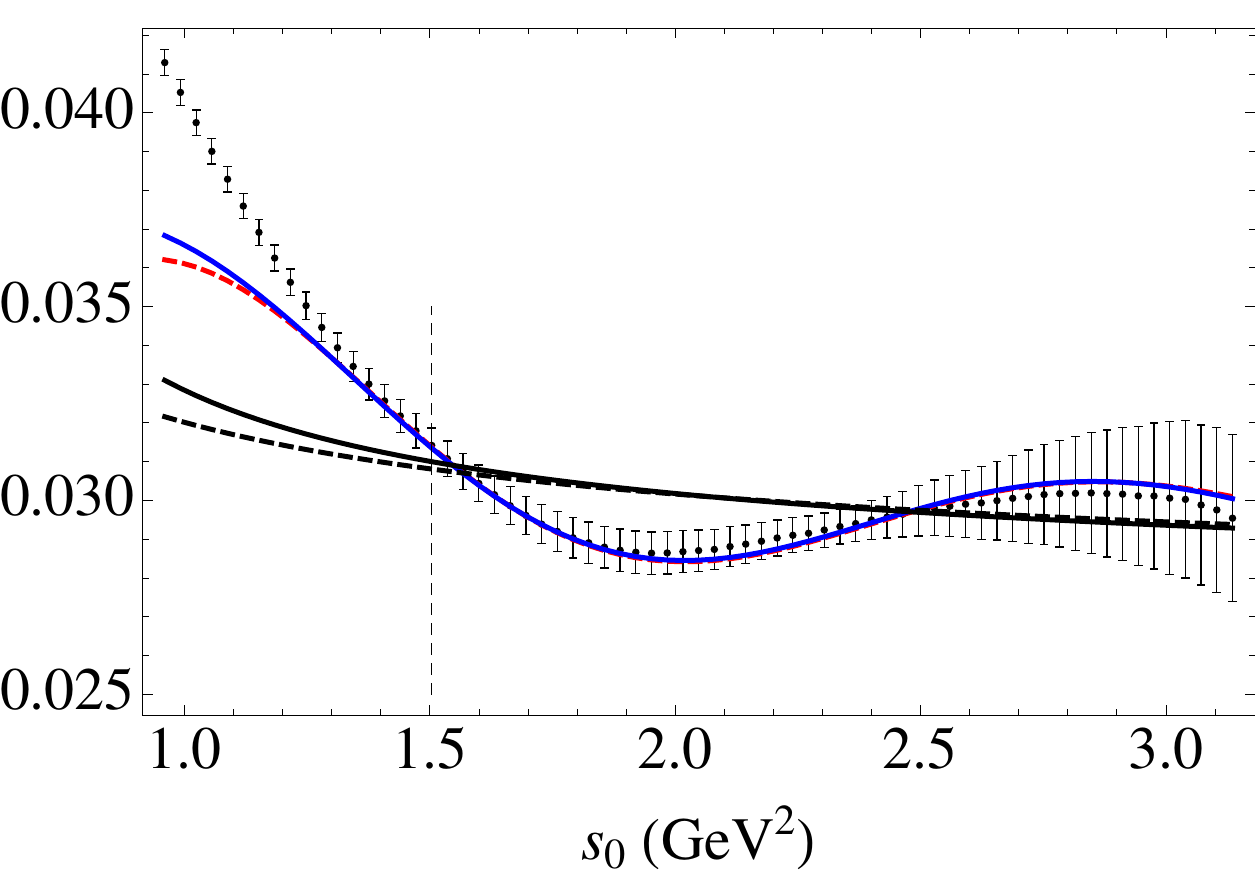}
\hspace{.1cm}
\includegraphics[width=2.9in]{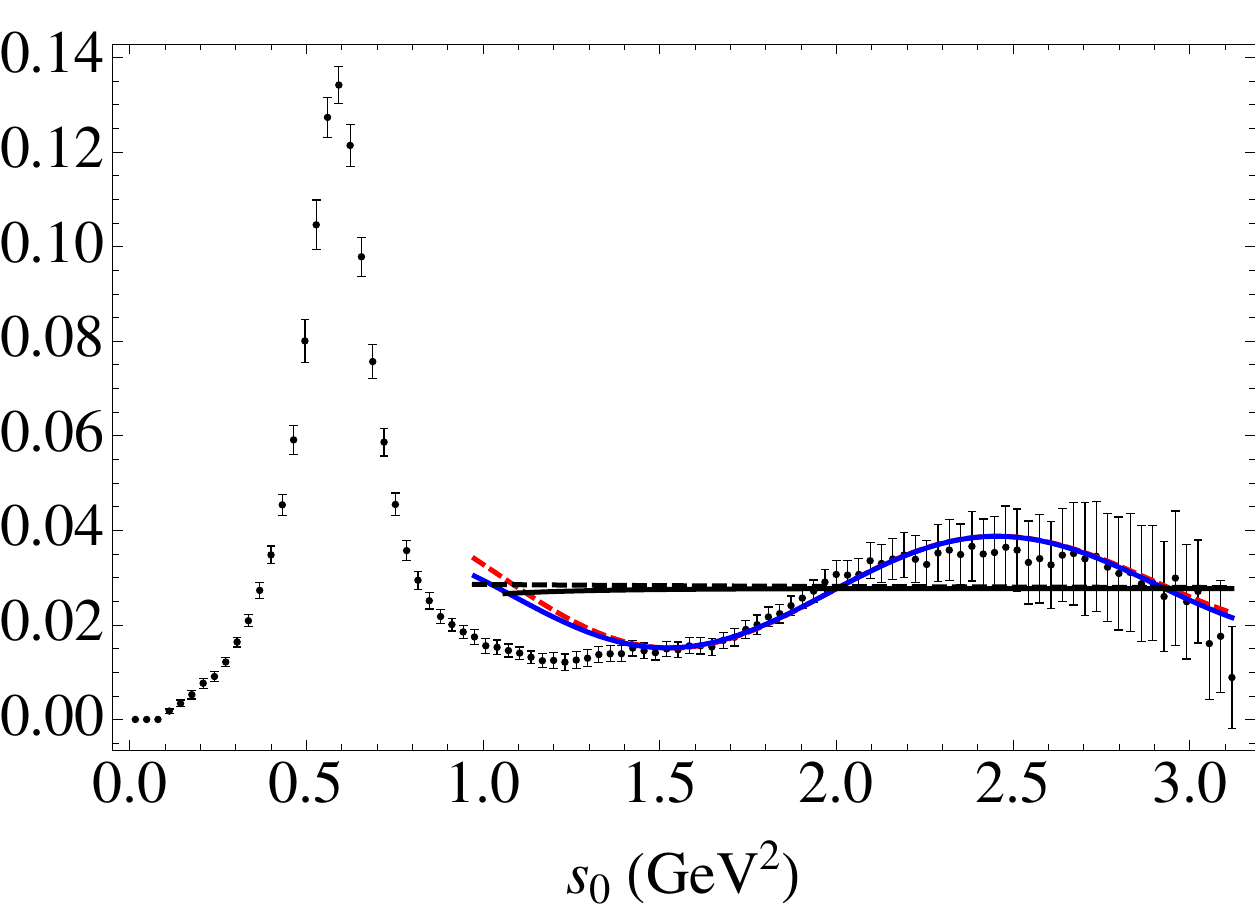}
\floatcaption{Vw0figure}{\it Left panel:
comparison of $I^{(\hw_0)}_{ex}(s_0)$
       and $I^{(\hw_0)}_{th}(s_0)$ for the $s_{min}=1.5\ {\rm GeV}^2$
       V-channel fits of Table~\ref{w0}.
   Right panel: comparison
of the theoretical spectral function resulting
from this fit with the experimental results.  CIPT fits are shown in red (dashed) and
FOPT in blue (solid).
The (much flatter) black curves represent the OPE parts of the fits.  The vertical dashed line
indicates the location of $s_{min}$.}
\vspace*{0ex}
\end{figure}

Figure~\ref{Vw0figure} shows the $\hat{w}_0$-FESR
fit for $s_{min}=1.5$~GeV$^2$ (left panel), and the corresponding 
theoretical curves for the spectral function in comparison with 
the (updated) experimental data (right panel).   
Agreement with data is good in the full fit window
$s_0\geq s_{min}= 1.5$~GeV$^2$. The black curves
show the OPE parts of the theoretical curves, \ie, the curves
obtained by removing the DV contributions from the blue and red
curves. Clearly, DVs are needed to give a good description
of the data for $I_{ex}^{(\hw_0)}$ and the spectral function itself.
We emphasize that the right panel of Fig.~\ref{Vw0figure} is not a fit;
only the moments $I^{(\hw_0)}(s_0)$ were used in the fits reported in
Table~\ref{w0}.

Fits with weight $\hw_0$ to the combined $V$ and $A$ channels are tabulated in Table~\ref{VAw0}, where again initial parameter estimates were
chosen close to the physical minima.   In view of our findings of Sec.~\ref{chi2}  for this case, we chose the thinning factor $n$ equal 
to 2.   We do not show plots of these fits, or of the corresponding spectral
functions, because they look very similar to those shown in Fig.~\ref{Vw0figure} and the corresponding figures in \paper.
The CIPT fit with $s_{min}=1.7$~GeV$^2$ appears to correspond to an unphysical solution of the type discussed in Sec.~\ref{chi2}; we did not find a physical solution in this case.

Following the same prescription as for Eq.~(\ref{asw0}), we obtain  for $\a_s$ the values
\begin{eqnarray}
\label{VAasw0}
\a_s(m_\t^2)&=&0.319\pm 0.015\pm 0.007\pm 0.005\qquad\mbox{(FOPT)}\ ,\\
\a_s(m_\t^2)&=&0.338\pm 0.021\pm 0.010\pm 0.004\qquad\mbox{(CIPT)}\nonumber
\end{eqnarray}
from the $V\&A$ fits.
The errors have the same meaning as in Eq.~(\ref{asw0}). The values in Eq.~(\ref{VAasw0}) 
are in good agreement with those of Eq.~(\ref{asw0}). It should, however,
be kept in mind that (i) the physical and unphysical minima of the
$\c^2$ function are close to degenerate for these fits (\seef\ Sec.~\ref{chi2})
and (ii) the averaged rescaling of the sum-of-residual-modes 
part of the OPAL spectral functions is considerably less reliable
for the $A$ channel than for the $V$ channel.

\begin{table}[h]
\begin{center}
\begin{tabular}{|c|c|c|c|c|c|c|c|}
\hline
$s_{min}$ & dof & $\c^2$/dof & $\a_s$ & $\d_{V/A}$ & $\g_{V/A}$ & $\a_{V/A}$ & $\b_{V/A}$ \\
\hline
1.3 & 49 & 0.58 & 0.327(12) & 3.71(55) & 0.38(39) & 0.24(51) & 3.00(30) \\
 &  & & & 1.62(86) & 1.66(55) & 2.48(77) & 3.60(46) \\
\hline
1.4 & 46 &  0.47 & 0.325(11) & 4.28(44) & 0.02(30) & -0.54(58) & 3.43(31) \\
 &  & & & 1.6(1.0) & 1.68(64) & 1.8(1.2) & 4.00(68) \\
\hline
1.5 & 43 & 0.53 & 0.312(15) & 3.90(71) & 0.29(46) & -1.02(85) & 3.64(45) \\
 & & & & 1.82(72) & 1.46(44) & -2.5(1.3) & 2.91(74) \\
\hline
1.6 & 40 & 0.40 & 0.320(13) & 4.23(57) & 0.05(35) & -0.79(70) & 3.55(37) \\
 & & & & 1.56(94) & 1.64(53) & 2.8(1.7) & 3.47(91) \\
\hline
1.7 & 37 & 0.54 & 0.312(17) & 3.7(1.4) & 0.42(78) & -0.9(1.0) & 3.60(51) \\
 & & & & 0.3(1.8) & 2.15(86) & -1.7(2.1) & 2.5(1.1) \\
\hline
\hline
1.3 & 49 & 0.61 & 0.348(18) & 3.38(51) & 0.58(38) & 0.30(54) & 2.93(32) \\
 &  & & & 1.95(78) & 1.48(50) & 2.51(83) & 3.61(49) \\
\hline
1.4 & 46 & 0.49  & 0.347(15) & 4.03(46) & 0.14(32) & -0.60(54) & 3.44(30) \\
 & &  & & 1.97(82) & 1.48(53) & 2.0(1.1) & 3.93(64) \\
\hline
1.5 & 43 & 0.54 & 0.328(21) & 3.69(77) & 0.39(51) & -1.08(83) & 3.66(45) \\
 & & & & 1.94(71) & 1.39(42) & -2.4(1.3) & 2.90(72) \\
\hline
1.6 & 42 & 0.40 & 0.339(17) & 4.09(61) & 0.12(38) & -0.90(67) & 3.59(35) \\
 & & & & 1.73(94) & 1.54(52) & 2.9(1.5) & 3.42(83) \\
\hline
1.7 & 37 & 0.47 & 0.294(17) & -0.9(2.8) & 3.3(1.6) & 3.1(2.3) & 8.2(1.2) \\
 & & & & 0.8(1.7) & 1.76(78) & -0.9(1.7) & 2.12(87) \\
\hline
\end{tabular}
\end{center}
%\vspace*{4ex}
\begin{quotation}
\floatcaption{VAw0}
{{\it Standard $\c^2$ fits to Eq.~(\ref{FESR}) for $w(s)=1$, combined V and A
channels. FOPT results are shown above the
double horizontal line, CIPT results below.
The first line for each $s_{min}$ gives the V DV parameters; the second line
the A ones.  Every second value of $s_0$ in the range above and
starting at $s_{min}$ is included in the fits.
Errors are standard $\c^2$ errors; $\g_{V/A}$ and $\b_{V/A}$ in {\rm GeV}$^{-2}$}}
\end{quotation}
\vspace*{-4ex}
\end{table}%
%\clearpage
%\pagebreak

%\newpage
\begin{boldmath}
\subsection{\label{moments} Fits with $\hw_{0,2,3}$}
\end{boldmath}
%%####%%
In this section, we report on simultaneous fits to moments with weights
$\hw_0$, $\hw_2$ and $\hw_3$, using the fit quality $Q^2$ of Eq.~(\ref{Q2}).
FOPT and CIPT results are shown for the $V$ channel in Table~\ref{w023}, for the same set of
$s_{min}$ values as before. 
To properly account for the cross-correlations between moments with 
different weights, errors and covariances are computed
through the linear-fluctuation analysis of \paper.
We do not show any plots based on these fits, as they 
look very similar to those in 
\paper.

\begin{table}[t]
\begin{center}
\vspace{0.1cm}
%\hspace{-1.4cm}
\begin{tabular}{|c|c|c|c|c|c|c|c|c|c|}
\hline
$s_{min}$ & dof & ${\cal Q}^2$/dof & $\a_s$ & $\d_V$ & $\g_V$ & $\a_V$ & $\b_V$ & $10^2C_{6,V}$ & $10^2C_{8,V}$ \\
\hline
1.3 & 167 & 0.42 & 0.307(13) & 2.68(75) & 1.10(51) & 0.29(80) & 2.89(47) & -0.51(35) & 0.68(66)\\
1.4 & 158 & 0.33 & 0.313(13) & 3.37(69) & 0.63(46) & -0.69(87) & 3.45(49) & -0.47(28) & 0.79(44)\\
1.5 & 149 & 0.33 & 0.315(14) & 3.74(60) & 0.40(39) & -0.9(1.0) & 3.55(57) & -0.45(28) & 0.80(45)\\
1.6 & 140 & 0.33 & 0.317(17) & 3.42(76) & 0.59(48) & -0.7(1.4) & 3.48(75) & -0.42(39) & 0.72(68)\\
1.7 & 131 & 0.33 & 0.318(19) & 4.26(73) & 0.14(39) & -0.8(1.3) & 3.53(68) & -0.46(38) & 0.86(61)\\
\hline
\hline
1.3 & 167 & 0.38 & 0.362(45) & 3.55(80) & 0.47(57) & 0.53(98) & 2.85(52) & -0.18(51) & 0.06(82)\\
1.4 & 158 & 0.30 & 0.349(30) & 3.85(66) & 0.30(44) & -0.3(1.0) & 3.28(55) & -0.40(33) & 0.53(55)\\
1.5 & 149 & 0.30 & 0.345(30) & 3.97(61) & 0.24(40) & -0.5(1.3) & 3.39(66) & -0.46(35) & 0.66(61)\\
1.6 & 140 & 0.31 & 0.347(42) & 3.71(71) & 0.38(47) & -0.4(1.8) & 3.33(92) & -0.42(55) & 0.6(1.0)\\
1.7 & 131 & 0.31 & 0.344(40) & 4.21(74) & 0.13(42) & -0.6(1.8) & 3.40(88) & -0.50(50) & 0.77(92)\\
\hline
\end{tabular}
\end{center}
%\vspace*{4ex}
\begin{quotation}
\floatcaption{w023}{{\it Fits to Eq.~(\ref{FESR}) with weights
$\hw_{0,2,3}$, V channel, using fit quality~(\ref{Q2}).
FOPT results are shown above the double horizontal line,
CIPT fits below. 
$\g_V$ and $\b_V$ in {\rm GeV}$^{-2}$, $C_{6,V}$ in {\rm GeV}$^6$ and
$C_{8,V}$ in {\rm GeV}$^8$.}}
\end{quotation}
\vspace*{-7ex}
\end{table}%

In this case, we have not carried out an investigation along the lines of 
Sec.~\ref{chi2}.  The reason is that we cannot compute a
fully correlated posterior probability distribution, 
and the interpretation of the probability distribution associated with 
$Q^2$ would be less clear. Our only reason for
considering these multiple-moment fits is to check that DVs in 
higher moments, and in particular the moment with the kinematic weight, 
can be described by our \ansatz, Eq.~(\ref{DVpar}).  We find that this is 
indeed the case.

The fit results, reported in Table~\ref{w023}, are in good agreement with those of Sec.~\ref{w=1}.  Following the same method as before, we obtain  for $\a_s$ the values
\begin{eqnarray}
\label{asw023}
\a_s(m_\t^2)&=&0.315\pm 0.014\pm 0.002\pm 0.007\qquad\mbox{(FOPT)}\ ,\\
\a_s(m_\t^2)&=&0.347\pm 0.030\pm 0.002\pm 0.005\qquad\mbox{(CIPT)}\ ,\nonumber
\end{eqnarray}
where again the errors have the same meaning as in Eq.~(\ref{asw0}).
For the CIPT case we
see that adding more moments has not improved the determination
and, in fact, has somewhat increased the total error. In view of
this observation, and the fact that the errors in Eqs.~(\ref{asw023}) and~(\ref{asw0})
were, in any case, produced using different minimizing functions, 
we stick with the standard $\c^2$ $V$-channel fit 
results of Eq.~(\ref{asw0}) as our central ones.

\begin{table}[t]
\begin{center}
%\hspace{-1.5cm}
\begin{tabular}{|c|c|c|c|c|c|c|c|c|c|}
\hline
$s_{min}$ & dof & ${\cal Q}^2$/dof & $\a_s$ & $\d_{V/A}$ & $\g_{V/A}$ & $\a_{V/A}$ & $\b_{V/A}$ & $10^2C_{6,V/A}$ & $10^2C_{8,V/A}$ \\
\hline
1.3 & 104 & 0.64 & 0.309(9) & 2.80(74) & 1.00(50) & -0.25(72) & 3.18(43) & -0.54(26) & 0.82(49) \\
& & & & 2.51(43) & 1.11(28) & 3.03(71) & 3.37(41) & 0.56(20) & -0.58(37) \\
\hline
1.4& 98 & 0.49 & 0.310(11) & 3.33(72) & 0.64(48) & -1.07(82) & 3.66(47) & -0.55(23) & 0.93(38) \\
&& & & 2.26(50) & 1.22(32) & -2.89(95) & 3.17(54) & 0.50(29) & -0.40(59) \\
\hline
1.5 & 92 & 0.48 & 0.312(11) & 3.70(62) & 0.40(41) & -1.23(95) & 3.75(53) & -0.52(22) & 0.92(36) \\
& & && 2.16(78) & 1.28(43) & -2.9(1.2) & 3.17(68) & 0.53(34) & -0.45(76) \\
\hline
1.6 & 86 & 0.38 & 0.292(14) & -0.8(3.0) & 3.2(1.8) & -0.9(1.8) & 7.0(1.0) & -1.14(18) & 2.10(32)\\
& && & 1.8(1.1) & 1.38(56) & -1.7(1.4) & 2.55(73) & -0.09(62) & 0.96(1.7) \\
\hline
1.7 & 80 & 0.44 & 0.312(17) & 3.78(89) & 0.36(50) & -1.1(1.4) & 3.69(74) & -0.52(37) & 0.91(66)\\
& && & -0.4(2.3) & 2.5(1.0) & -1.6(3.2) & 2.5(1.7) & 0.27(78) & 0.5(2.4) \\
\hline
\hline
1.3 & 104 & 0.53 & 0.346(18) & 3.45(58) & 0.54(41) & 0.02(70) & 3.09(40) & -0.38(27) & 0.43(46) \\
& && & 1.97(69) & 1.46(44) & 2.33(76) & 3.73(44) & 0.71(22) & -1.02(43) \\
\hline
1.4 & 98 & 0.43 & 0.339(17) & 3.75(57) & 0.34(39) & -0.72(83) & 3.49(46) & -0.51(22) & 0.73(38) \\
& && & 2.03(60) & 1.39(39) & 2.8(1.0) & 3.48(59) & 0.59(26) & -0.76(55) \\
\hline
1.5 &92 & 0.43 & 0.337(17) & 3.89(53) & 0.26(36) & -0.95(98) & 3.60(53) & -0.55(23) & 0.82(41) \\
& && & 2.13(78) & 1.34(45) & 2.8(1.4) & 3.47(74) & 0.58(31) & -0.74(68) \\
\hline
1.6 & 86 & 0.44 & 0.335(23) & 3.56(77) & 0.45(48) & -0.9(1.4) & 3.57(73) & -0.55(34) & 0.79(65)\\
& && & 1.7(1.1) & 1.51(58) & 3.1(1.8) & 3.31(99) & 0.50(46) & -0.5(1.1) \\
\hline
1.7 & 80 & 0.42 & 0.332(30) & 3.79(84) & 0.33(47) & -1.0(1.7) & 3.61(85) & -0.58(43) & 0.88(84)\\
& & & & -0.3(2.4) & 2.14(1.0) & -2.1(3.5) & 2.7(1.9) & 0.30(85) & 0.2(2.5) \\
\hline
\end{tabular}
\end{center}
%\vspace*{4ex}
\begin{quotation}
\floatcaption{VAw023}{{\it Fits to Eq.~(\ref{FESR}) with
weights $\hw_{0,2,3}$, combined $V$ and
$A$ channels, using fit quality~(\ref{Q2}). FOPT results are shown above
the double horizontal line, CIPT fits below.
$\g_{V/A}$ and $\b_{V/A}$ in {\rm GeV}$^{-2}$, $C_{6,V}$ and $C_{6,A}$ in
{\rm GeV}$^6$ and
$C_{8,V}$ and $C_{8,A}$ in {\rm GeV}$^8$.
The first line for each $s_{min}$ gives the $V$ channel DV and OPE parameters;
the second line the $A$ channel ones.
Every third value of $s_0$ in the range above and starting at $s_{min}$
is included in the fits.}}
\end{quotation}
\vspace*{-4ex}
\end{table}%
%\clearpage

In Table~\ref{VAw023} we show similar results for the $V\&A$ analysis.
In this case, we found the most stable results with the thinning factor $n=3$,
\ie, thinning out the moments $I^{(\hw_{0,2,3})}_{\rm ex}(s_0)$ by a factor three.
Even so, we did not find a physical minimum for the FOPT case with $s_{min}=1.6$~GeV$^2$, as can be seen from the table.   The distinction
between the physical minima we found at lower values of $s_{min}$ and the
unphysical minima at higher values of $s_{min}$ is very clear from the
values of the DV parameters.\footnote{The value of $\c^2$ is always smaller at the unphysical minimum in these particular fits.}  In particular $\d_V$ and $\g_V$ both 
differ by a large amount between physical and unphysical solutions, much as shown in the simpler case displayed in Fig.~\ref{Vw0proj}.  Averaging only the FOPT fits at $s_{min}=1.4$ and $1.5$~GeV$^2$, and averaging as before the CIPT fits at $s_{min}=1.4$, $1.5$ and $1.6$~GeV$^2$, we find for $\a_s(m^2_\t)$
\begin{eqnarray}
\label{VAasw023}
\a_s(m_\t^2)&=&0.311\pm 0.011\pm 0.002\pm 0.007\qquad\mbox{(FOPT)}\ ,\\
\a_s(m_\t^2)&=&0.337\pm 0.017\pm 0.002\pm 0.005\qquad\mbox{(CIPT)}\ ,\nonumber
\end{eqnarray}
with errors again as in Eq.~(\ref{asw0}).

\vspace*{5ex}
%\newpage
\section{\label{summary} Summary of results}
%%####%%
Through a more detailed statistical study of the data than we carried out in
\paper, we showed that our fits of the OPAL data sometimes allow for different local mimina of the $\c^2$ function, \seef\ Sec.~\ref{chi2}.   These
solutions are most clearly distinguished by the values of the DV parameters
$\d_V$ and $\g_V$, and in the introduction to Sec.~\ref{fits} we argued that the solutions with large values of $\d_V$ (of order 4) and small values of $\g_V$ 
(of order 0.3) should be considered as physical, while the other minima, which
always have small values of $\d_V$ (negative, in fact), and large
values of $\g_V$ (typically of order 3) should not be considered physical.

As in \paper, we have 
used both Weinberg sum rules \cite{SW} and the sum rule for the 
electro-magnetic pion mass difference \cite{EMpion}
to test our fit results. All three sum rules are well satisfied, at a level 
of precision similar to that found in \paper. We have also confirmed, 
again as in \paper, that our theoretical description of 
$R^{(1+0)}_{V+A,ud}(s_0)$ (Eq.~(\ref{Rt})) agrees, within errors,
with data for $s_0$ down to below $1.5$~GeV$^2$.

%\newpage
\begin{boldmath}
\subsection{\label{alphas} The value of $\alpha_s(m_\tau^2)$}
\end{boldmath}
%%####%%
We will choose the values of the strong coupling at the $\t$ mass obtained from the $V$-channel fit of $I^{(\hw_0)}_{\rm ex}$ as our best values.
Our reasoning for doing so is twofold.   First, while the simultaneous
fits to multiple moments are in good agreement with this simple fit, no
standard $\c^2$ fit is possible in this case.   While we believe that the
error estimates based on linear fluctuation analysis are reasonable, it is
less clear how they should be interpreted than those obtained
from a standard $\c^2$ analysis.\footnote{Of course, we have found that the
posterior probability distribution has a complicated behavior, so that physical
input is required to decide which local minimum is physical, as discussed in
detail in Secs.~\ref{chi2} and \ref{fits}.}  Second, including also the $A$
channel makes the fits more complicated, because of the larger number of
parameters. 
This effect is compounded by the much larger, and hence less certainly
reliable, approximate rescaling that must be applied to 
the residual distribution in the $A$ channel.
In addition, we note that the only feature visible in the $A$ channel is 
the $a_1$ resonance, and it is not clear whether the \ansatz\ we use 
to parameterize the DV part of the spectral function can be expected to 
apply to this resonance, even if we assume that the \ansatz\ works well 
for higher resonances in each channel. Finally, related to this, we note that the typical values of the DV parameters
we find in the axial channel satisfy the expectation of Eq.~(\ref{parameters}) less well.

We thus find our best values for the strong coupling from Eq.~(\ref{asw0}):
\begin{eqnarray}
\label{asw0again}
\a_s(m_\t^2)&=&0.325\pm 0.018\qquad(\overline{MS},\ n_f=3,\ \mbox{FOPT})\ ,\\
\a_s(m_\t^2)&=&0.347\pm 0.025\qquad(\overline{MS},\ n_f=3,\ \mbox{CIPT})\ ,\nonumber
\end{eqnarray}
where  we added the errors in Eq.~(\ref{asw0}) in quadrature.

Running these values up to the $Z$ mass $M_Z$ \cite{CKS} yields\footnote{We
evolved $\a_s$ to the $Z$ mass in the same way as was done in Ref.~\cite{BJ}.
Uncertainties in the running, associated with the use of
4-loop truncated $\beta$ functions, uncertainties in the charm and
bottom masses, and the choice of the $n_f=3\rightarrow n_f=4$ and
$n_f=4\rightarrow n_f=5$ matching thresholds are negligible on the scale
of the quoted errors.}
\begin{eqnarray}
\label{finalZ}
\a_s(M_Z^2)&=&0.1191\pm 0.0022\qquad(\overline{MS},\ n_f=5,\ \mbox{FOPT})\ , \\
\a_s(M_Z^2)&=&0.1216\pm 0.0027\qquad(\overline{MS},\ n_f=5,\ \mbox{CIPT})\ ,\nonumber
\end{eqnarray}
where we symmetrized the resulting slightly asymmetric errors.

%\newpage
\subsection{\label{condensates} Non-perturbative results}
%%####%%
In \paper\ we estimated the relative deviation of the values found for the
dimension-6 condensates from those given by vacuum-saturation approximation. To this end,
these condensates, parametrized by $C_{6,V/A}$, are expressed in terms of
the quantities $\r_1$ and $\r_5$ by
\begin{equation}
\label{C6VA}
C_{6,V/A} \,=\, \frac{32}{81}\,\pi^2 a_s\,
\langle\bar qq\rangle^2 \!\left(\!\!\begin{array}{c} 2\,\rho_1 -
9\,\rho_5 \\ 11\,\rho_1 \end{array}\!\!\right)\! \ .
\end{equation}
Vacuum saturation values for $C_{6,V/A}$ then correspond to $\r_1=\r_5=1$.
Performing the analogous analysis for the updated OPAL spectral functions,
we find (employing
$\langle\bar qq\rangle(m_\tau^2)=-\,(272\,{\rm MeV})^3$ \cite{jam02})
\begin{eqnarray}
\label{rho1rho5num}
\rho_1 &\!\!=\!\!& \,3.1 \pm 2.0 \,, \quad
\rho_5 \,=\, 4.4 \pm 1.4 \qquad \mbox{(FOPT)} \ , \\
%\vbox{\vskip 6mm}
\rho_1 &\!\!=\!\!& \,3.1 \pm 1.6 \,, \quad
\rho_5 \,=\, 4.3 \pm 1.3 \qquad \mbox{(CIPT)} \ ,\nonumber
\end{eqnarray}
using as representative values $C_{6,V/A}$ and $\a_s(m^2_\t)$ of
Table~\ref{VAw023} for $s_{min}=1.5\,{\rm GeV}^2$.\footnote{We neglect the
errors on $\a_s$ and $\langle\bar qq\rangle$.} As in \paper, the values of
$\r_1$ and $\r_5$ are insensitive to the perturbative resummation scheme. We note
that $\r_1$ changes sign relative to the central value found in \paper, but
also that, given the large uncertainties, there is no inconsistency between
our earlier fits and those presented here.

Analyses of the strong coupling from $\tau$ decays are sometimes based
on the ratio, $R_{V+A}^\tau$, of the total
inclusive non-strange branching fraction to the electron
branching fraction $B_e$ \cite{BNP},
\begin{eqnarray}
\label{Rtau}
R^{\t}_{V+A}=N_c S_{\mathrm{EW}}|V_{ud}|^2 \left(1+ \d_P+\d_{NP}\right)\ ,
\end{eqnarray}
where $\d_P$ stands for the perturbative, and $\d_{NP}$ stands for the non-perturbative contributions beyond the parton model. 
Determining $\d_{P}$, and hence $\a_s$, from $R^\t_{V+A}$
of course requires input for $\d_{NP}$. In the past, shortcomings
in the methods used to obtain this input have led to a significant
underestimate of the corresponding uncertainties. Analyses (such as
those of Refs.~\cite{OPAL,ALEPH}) including additional higher-degree-weight
FESRs, for example, were forced to assume that $D>8$ contributions
could be neglected for all additional FESRs.\footnote{This assumption has since been
tested (and found to be poorly satisfied) by comparing the
$s_0$-dependence of the fitted theory side to that of the experimental
data side of the various additional FESRs \cite{MY}.}
 Reference~\cite{MY} avoided this
problem, but, being unable to fit all required $D\le 8$ OPE parameters
using an $s_0$ window within which neglect of integrated DVs was
self-consistent, was forced to rely on external input for the gluon
condensate, the renormalon ambiguity of which makes this external input
potentially problematic. As shown in \paper\ and Ref.~\cite{DV7phipsi}, it is
not possible to avoid these problems without considering lower $s_0$
and FESRs for which integrated DVs are not negligible in the full
$s_0$-fitting window employed. The framework presented in \paper\ and in the present article is the first to
allow for a reliable estimate of $\d_{NP}$ from such an analysis, and hence
to bring these  systematic issues on the theory side under control. Expressing
the $D=6$, $8$ OPE terms, as well as the DV contributions to
$\d_{NP}$, in terms of $\d^{(6)}$, $\d^{(8)}$ and
$\d^{DV}$, respectively, we obtain from our fits with $s_{min}=1.5$
GeV$^2$,
\begin{eqnarray}
\label{delta68DV}
\delta^{(6)} &\!\!=\!\!& (\phantom{-}\,0.0 \pm 1.9)\cdot 10^{-2} \,,\quad
\delta^{(8)} \,=\, (-\,3.7 \pm 7.6)\cdot 10^{-3} \,, \nonumber \\
\delta^{{\rm DV}} &\!\!=\!\!& (-\,0.1 \pm 1.0)\cdot 10^{-3}
\qquad \mbox{(FOPT)} \ , \\
\vbox{\vskip 6mm}
\delta^{(6)} &\!\!=\!\!& (-\,0.1 \pm 1.8)\cdot 10^{-2} \,, \quad
\delta^{(8)} \,=\, (-\,0.6 \pm 7.6)\cdot 10^{-3} \,, \nonumber \\
\delta^{{\rm DV}} &\!\!=\!\!& (-\,0.6 \pm 1.4)\cdot 10^{-3}
\qquad \mbox{(CIPT)} \ . \nonumber
\end{eqnarray}
In the case of FOPT, the corresponding correlation matrix is found to be:
\vspace*{2ex}
\begin{table}[h]
\begin{center}
\begin{tabular}{|c|ccc|}
\hline
 & $\d^{(6)}$ & $\d^{(8)}$ & $\d^{{\rm DV}}$ \\
\hline
$\d^{(6)}$      &  1    &  -0.98 & 0.59 \\
$\d^{(8)}$      &  -0.98 &  1    & -0.54 \\
$\d^{{\rm DV}}$ & 0.59 & -0.54 &  1    \\
\hline
\end{tabular}
\end{center}
\begin{quotation}
\floatcaption{deltacorr}{{\it Correlation matrix for the quantities
$\delta^{(6)}$, $\delta^{(8)}$ and $\delta^{{\rm DV}}$ of
Eq.~(\ref{delta68DV}) (FOPT).}}
\end{quotation}
\vspace*{-4ex}
\end{table}%

While individually for the $V$ and $A$ channels the hierarchy of
the non-perturbative terms is such that dimension-6 is the largest, dimension-8
smaller, and the DV contribution the smallest, due to strong cancellations in
the $D=6$ and DV contributions for the sum $V+A$, the $D=8$ contribution turns out to be dominant.
However, in view of the large uncertainties, it is impossible to conclude that
these cancellations will also persist once more precise data are available.

Combining the OPE contributions as well as the DV term of Eq.~(\ref{delta68DV})
including correlations, the total non-perturbative contribution to $R^{\t}_{V+A}$ turns out to be
\begin{eqnarray}
\label{deltaNP}
\delta^{{\rm NP}} &\!\!=\!\!& (-\,0.4 \pm 1.2)\cdot 10^{-2}
\qquad \mbox{(FOPT)} \ , \\
%\vbox{\vskip 6mm}
\delta^{{\rm NP}} &\!\!=\!\!& (-\,0.2 \pm 1.2)\cdot 10^{-2}
\qquad \mbox{(CIPT)} \ . \nonumber
\end{eqnarray}
These estimates, despite having errors larger than those quoted
previously in the literature, must be considered more reliable,
as they are the only ones based on an analysis which deals explicitly with the
theoretical systematic issues discussed above.

  Care must be taken in drawing conclusions from the results of
Eq.~(\ref{delta68DV}). While the results {\it do} establish that integrated DV
contributions to $R^\t_{V+A}$ are small, it does {\it not} follow,
as repeatedly assumed in the literature,{\footnote{For a recent review,
see \eg\ Ref.~\cite{Pich11}.}} that DVs can be neglected in the determination of
$\a_s$ from hadronic $\t$ decay data. The reason is that, even if
one restricts attention to only the quantity $R^\t_{V+A}$, one still needs to
determine the $D=6$ and  $D=8$ contributions to $\d_{NP}$. 
This cannot be done in a controlled manner without including
values of $s_0$ significantly lower than $m_\t^2$ and weights for which integrated DV
contributions are certainly not negligible (\seef\ \paper\ and Ref.~\cite{DV7phipsi}). Many
values of $\d_P$ in the literature
 have been obtained using values of $\d_{NP}$ taken
from analyses with the limitations noted above. In view of the results
given in Eq.~(\ref{deltaNP}), the errors on such estimates of $\d_P$
are evidently underestimated, often by a significant amount. Only improved data will allow these errors to be further reduced.

\begin{figure}[t]
%\vspace*{4ex}
\begin{center}
\includegraphics*[width=7cm]{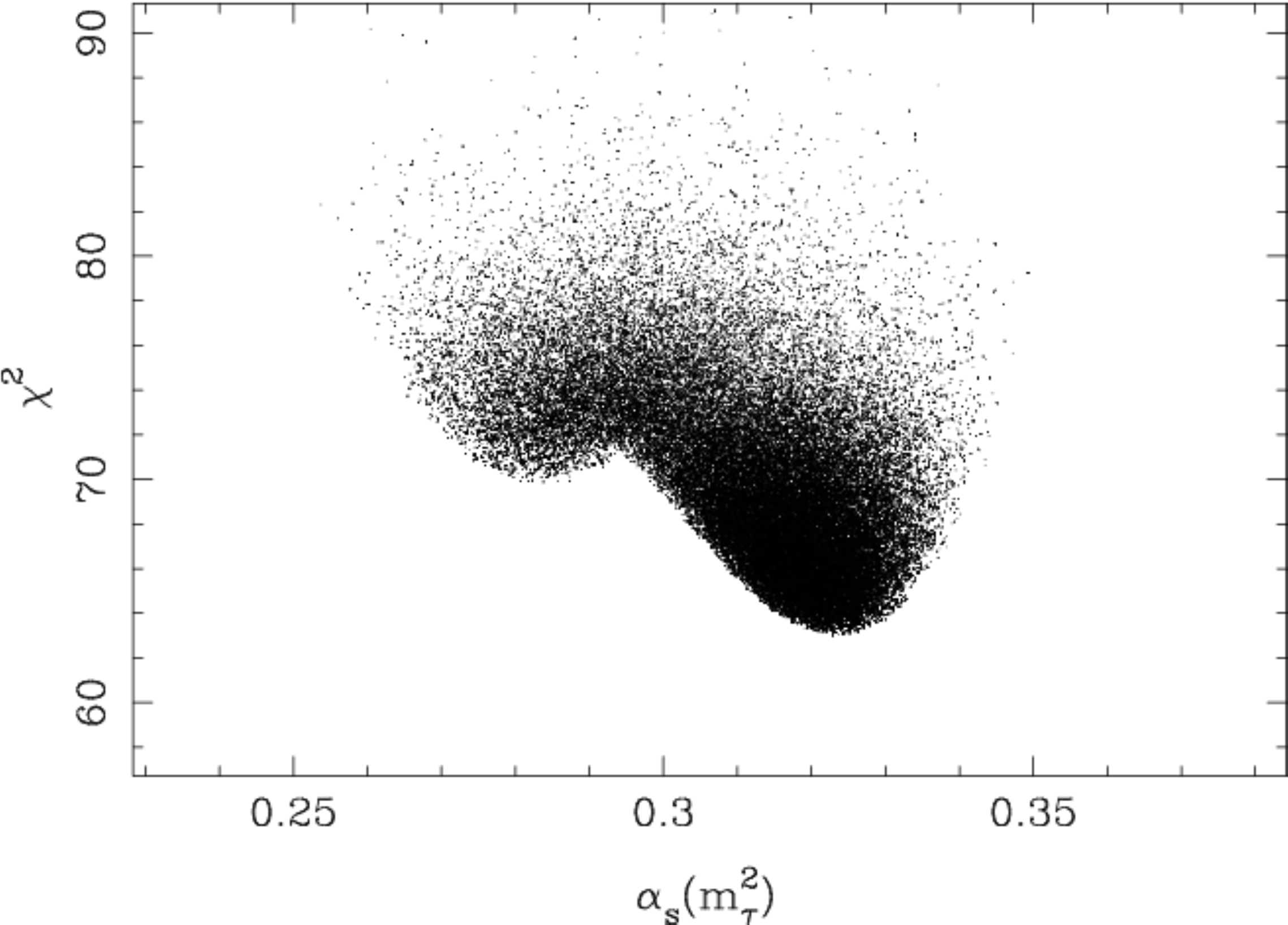}
\hspace{.1cm}
\includegraphics*[width=7cm]{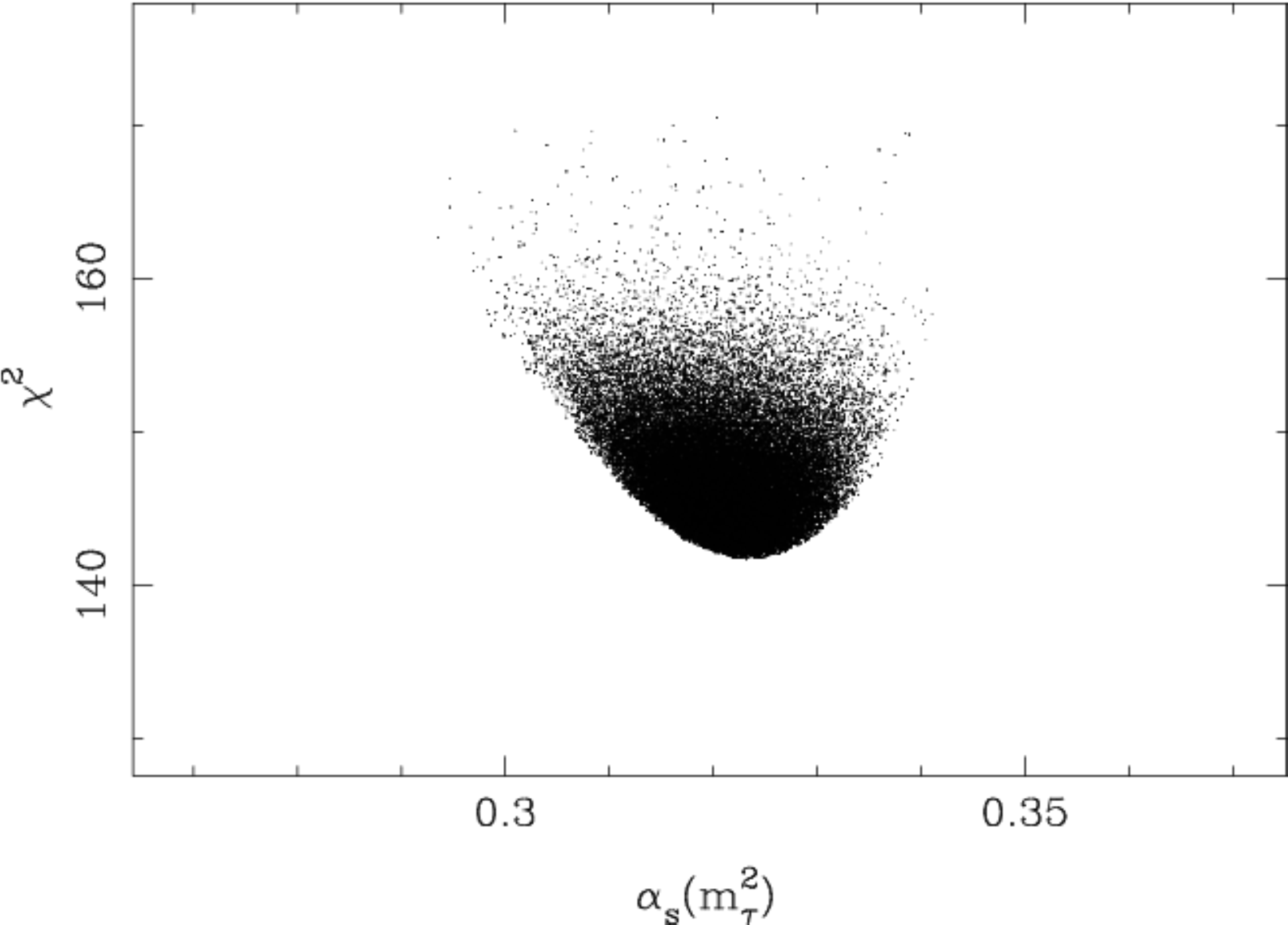}
\end{center}
\begin{quotation}
\floatcaption{futurechi2}%
{{\it $\a_s(m^2_\t)$ versus $\c^2$,  covariance matrix reduced by factor 4 (left panel) and factor 9 (right panel),
$V$ channel, with $s_{min}=1.5$~{\rm GeV}$^2$.}}
\end{quotation}
\vspace*{-4ex}
\end{figure}
\begin{figure}[h]
%\vspace*{4ex}
\begin{center}
\includegraphics*[width=7cm]{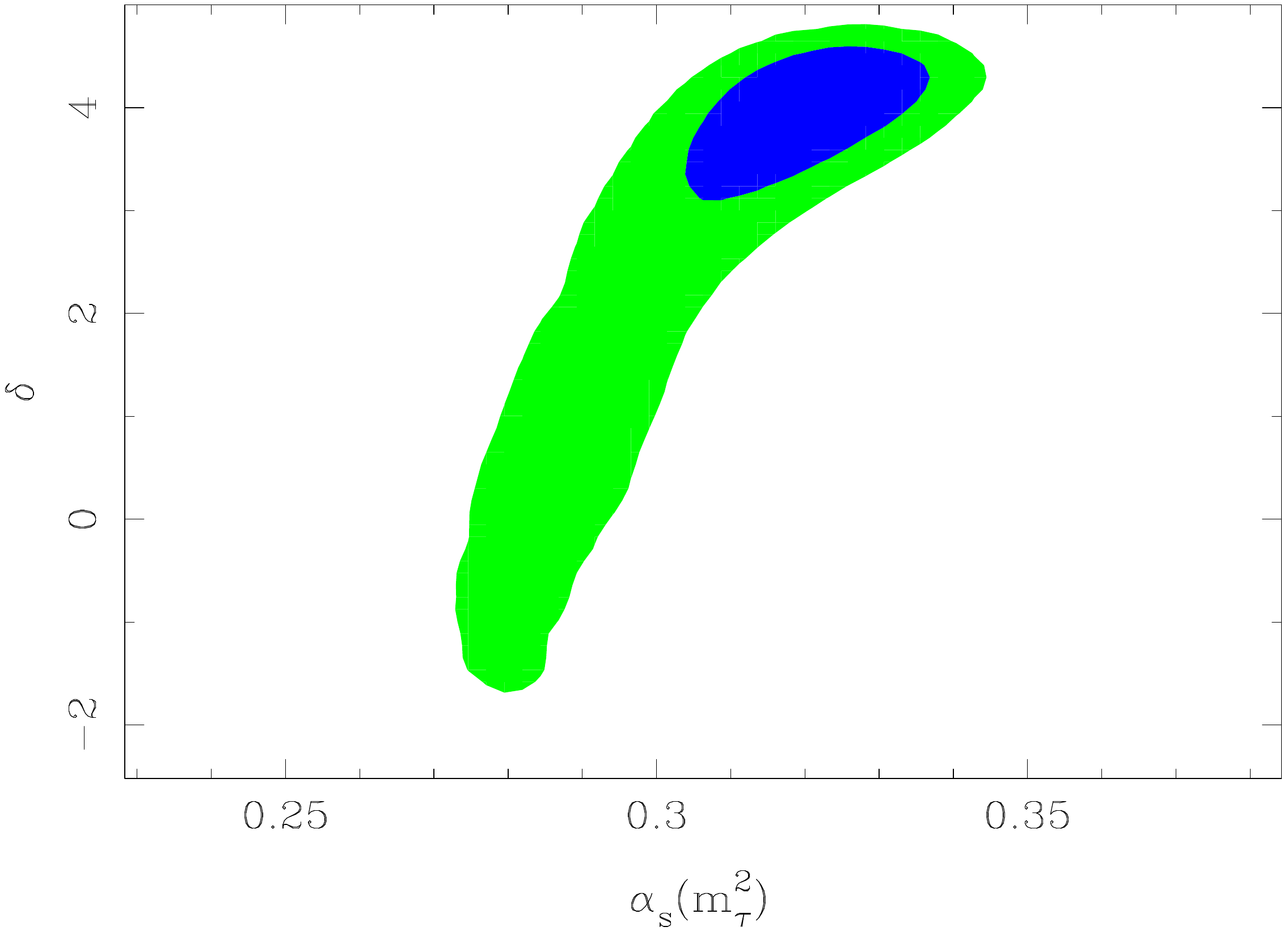}
\hspace{.1cm}
\includegraphics*[width=7cm]{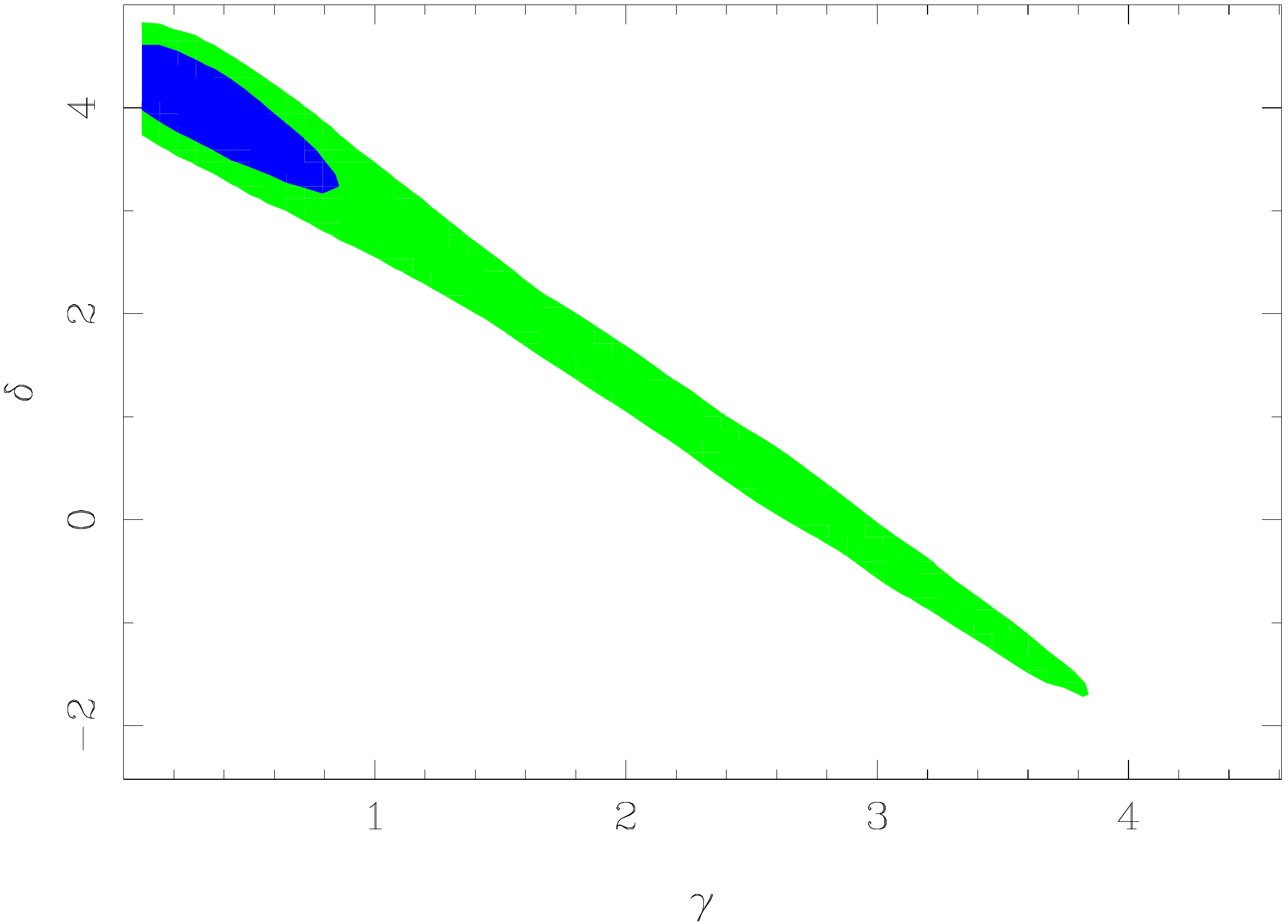}
\vskip0.5cm
\includegraphics*[width=7cm]{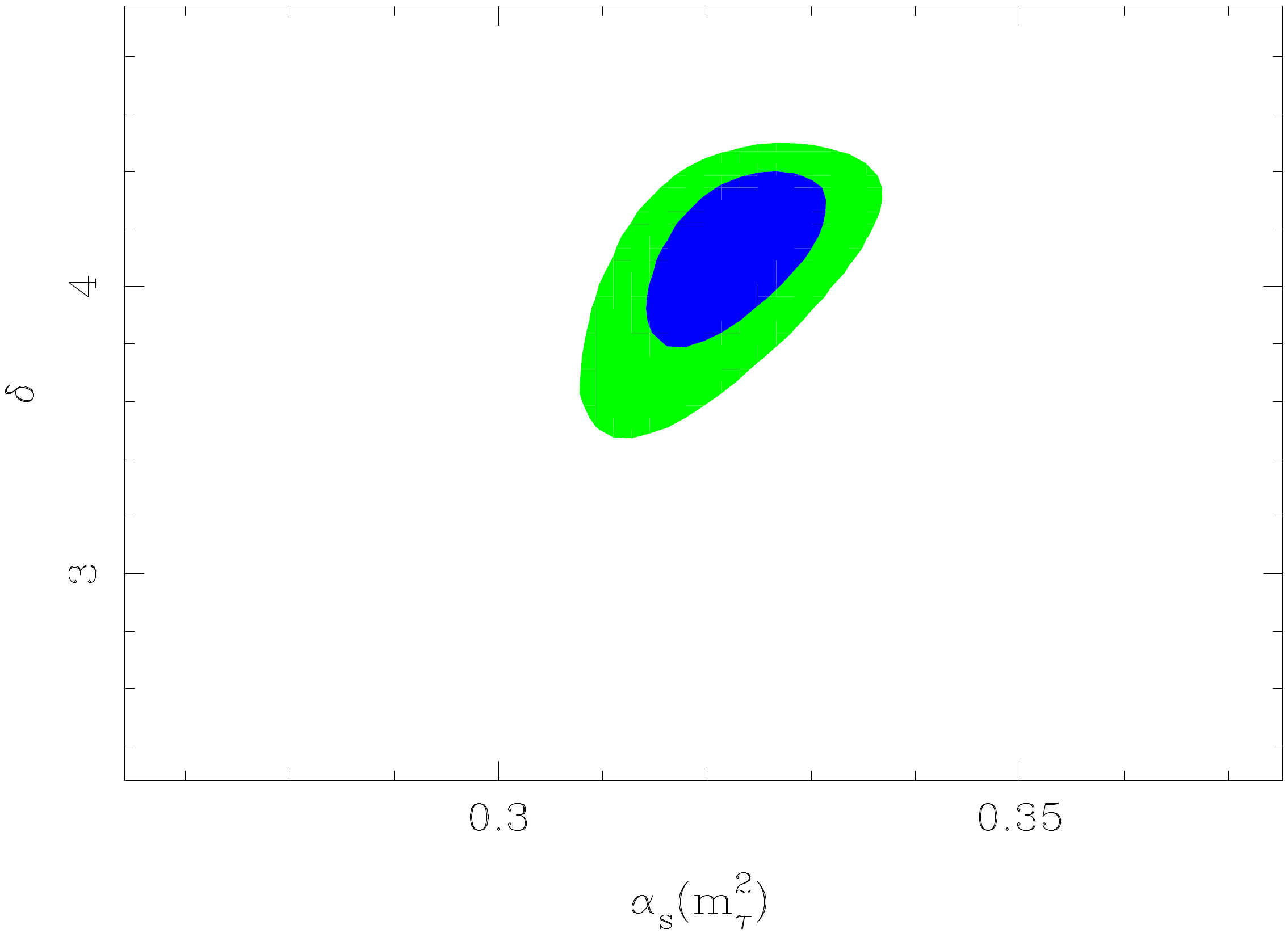}
\hspace{.1cm}
\includegraphics*[width=7cm]{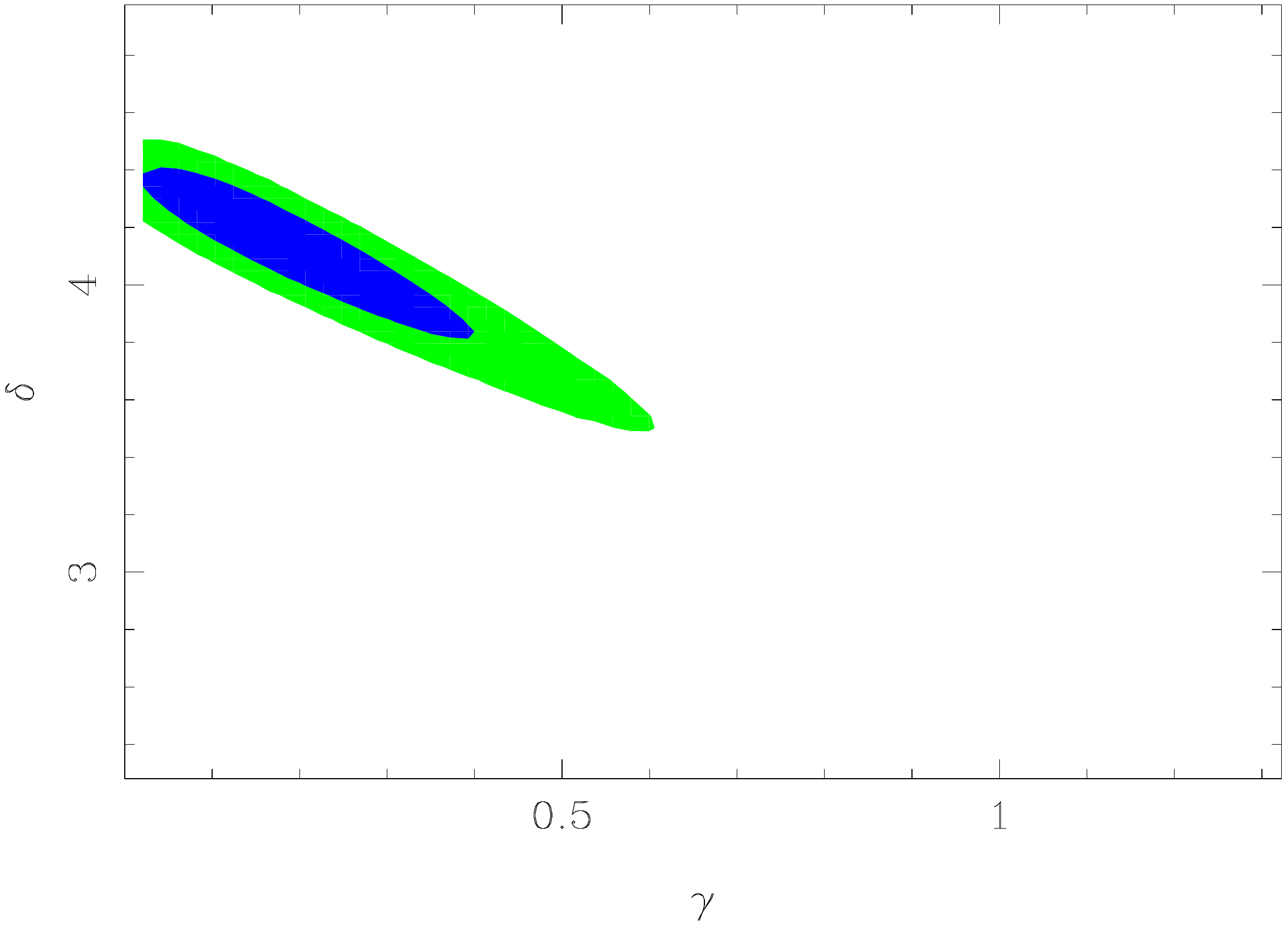}
\end{center}
\begin{quotation}
\floatcaption{futurecontour}%
{{\it  Two-dimensional contour plots showing $\a_s(m^2_\t)$ versus $\d_V$ and
$\g_V$ versus $\d_V$.
Top: covariance matrix reduced by factor 4; bottom: covariance matrix reduced by factor 9.
Left panel: projection onto $\a_s(m^2_\t)-\d_V$ plane; right panel: projection onto $\g_V-\d_V$ plane.  $V$ channel, $s_{min}=1.5\ {\rm GeV}^2$.  Blue (darker) areas and green (lighter) areas contain
68\%, respectively, 95\% of the distribution.}}
\end{quotation}
\vspace*{-4ex}
\end{figure}
%%%%%%%%%%%%%%%%%%%

%\newpage
\section{\label{future} Future perspectives}
%%####%%
%%%%%%%%%%%%%%%%%%%
In this section, we speculate on possible improvements relative to the 
results presented in Sec.~\ref{chi2} if data with significantly smaller 
errors were to become available.  As mentioned before, in principle 
such data can be extracted from the BaBar and Belle experimental results, 
and it is not unlikely that such an analysis would lead to hadronic spectral 
functions with errors about 2 or 3 times smaller than those of the OPAL data,
especially in the upper part of the kinematic region, where OPAL 
statistics are low.

Therefore, in Figs.~\ref{futurechi2} and~\ref{futurecontour}, we consider how the results shown in Figs.~\ref{as-chi2} and~\ref{Vw0proj} would change if we used the same central values for the $V$ spectral function as used in Sec.~\ref{chi2},
but with a covariance matrix scaled by a factor $1/4$ or $1/9$.   
We emphasize that this is just a speculative exercise.   For instance, given that the values of the $\c^2$ per degree of freedom in Table~\ref{w0} are of order one, fits with these scaled covariance matrices would give rise to fits yielding the same central values, but with poor values of the $\c^2$ per degree of freedom.    This observation reflects just the fact that the data can, of course, not be improved by rescaling the covariance matrix, simply because fluctuations in the actual data correspond to the size of the actual covariance matrix.

However, it is of some interest to see what would happen to the properties of the conditional probability distribution we explored in Sec.~\ref{chi2}.  
The new figures all show that the unphysical minimum of Fig.~\ref{as-chi2}
disappears as a function of the rescaling factor, while leaving the physical minimum in place.   We interpret this as evidence that better data, \ie, data with smaller errors, may help resolve the problem that with current data the parameter $\d_V$ (and therefore, because of the strong correlations, the
parameters $\g_V$ and $\a_s(m^2_\t)$) cannot be reliably determined without external considerations (\seef\ Sec.~\ref{fits}).   Of course, this exercise assumes that better data would be equally well described by our theoretical parameterization of the spectral-function moments we consider in this article.

%\newpage
\section{\label{conclusion} Conclusion}
%%####%%
In this article, we continued our analysis of hadronic $\t$ decays.   The main goal is the precision determination of the strong coupling at the $\t$ mass, $\a_s(m^2_\t)$, with good control not only over statistical, but also over systematic errors.
In our previous article, \paper, we presented a new framework for such an 
analysis, in which non-perturbative contributions to the non-strange vector and axial hadronic $\t$ decays, both from the operator product expansion and from violations of quark-hadron duality, can be quantitatively estimated.   Since complete spectral functions for these channels are available from experiment, the energy dependence of these effects can be taken into account.   The results of Ref.~\cite{MY}, \paper, and the present article show unambiguously that it is imperative to take this energy dependence into account in order to arrive at a fully consistent understanding of non-perturbative effects.   This requires the use of a model to parameterize duality violations.   
We emphasize that a quantitative approach to duality violations cannot be avoided, simply because they are clearly present in the spectral-function data.
The assumption that
duality violations are negligible, while perhaps reasonable in the past, is no
longer acceptable given the current-claimed level of precision.\footnote{All results reviewed in Ref.~\cite{Pich11} claim an error $\leqx 0.015$
on $\a_s(m_\t^2)$.}
 
The specific aim of the present article is two-fold.   First, the analysis of \paper\ was based on the 1998 OPAL spectral-function data.   The construction of these spectral functions included the use of the then available values for the
branching fractions for the most important exclusive modes.   Since these branching fractions are now more precisely known, it is possible to update the central values for the spectral functions as well as the corresponding diagonal errors.\footnote{A partial update of the full covariance matrix is also possible, as explained in the Appendix.}   This update was carried out in Sec.~\ref{OPALupdate}, and the results were subsequently used in our fits to the data.

Our second aim was a more detailed investigation of the quality of the fits that go into our analysis.   Since these are non-linear, multi-parameter fits, they are
of considerable complexity.   The use
of a Markov-chain Monte Carlo program made it possible to investigate the
full posterior probability distribution underlying our most important fits. 
This allowed us to delineate the landscape in parameter space in more detail than through simple minimization.  This investigation was carried out in
Sec.~\ref{chi2}. 

Let us summarize what we learned from our new analysis. First, as the
reader will note, our new
results for the value of $\a_s(m^2_\t)$, contained in Eq.~(\ref{asw0again}), 
are very close to the OPAL results of Ref.~\cite{OPAL},
but with somewhat larger errors. 
Since the two sets of results correspond to sets of 
data with different normalizations, the near-equality of
central values is, in fact, purely accidental; as shown in
\paper, an analysis of the same data as that used by OPAL
leads instead to significantly smaller values of $\a_s(m^2_\t)$.
A related observation applies to the errors. 
The errors found in Ref.~\cite{OPAL} are smaller simply 
because systematic effects associated with the operator 
product expansion and duality violations were not considered in 
that analysis. In our opinion, the same observation applies to essentially 
all determinations of $\a_s(m^2_\t)$ from hadronic $\t$ decays preceding 
the framework presented in \paper.

Second, while we argued in \paper\ and here that duality violations cannot 
be reliably left out from a quantitative analysis of the spectral functions 
below the $\t$ mass, it turns out that the multi-parameter fits thus needed to determine all parameters are at the edge of what is possible with currently 
available data.  This is demonstrated in Figs.~\ref{as-chi2},~\ref{Vw0proj} 
and~\ref{as-chi2VA}, which show that the probability distributions 
underlying our fits may have several minima, which together span a 
range of $\a_s$ values of about $0.27-0.34$.\footnote{This is for 
FOPT; for CIPT the range is shifted by about $0.02$.}  Therefore, physical
arguments, given in Sec.~\ref{fits}, are needed in order to narrow down the
error on $\a_s(m^2_\t)$, and our result~(\ref{asw0again}) is obtained with 
the help of these arguments.

Given this state of affairs, we believe that it would be very interesting 
to apply our analysis to  data with much better statistics, which are in 
principle available from the BaBar and Belle experiments.  If the non-strange 
spectral functions that can be extracted from these data would be made 
available, this would allow us to put our analysis framework to a much 
more stringent test. This was demonstrated quantitatively in Sec.~\ref{future}, 
where it was shown that with much reduced statistical errors one may expect 
to resolve the ambiguities present in the probability distribution 
constructed from the OPAL data.

Of course, at present we do not know what the outcome of such an 
investigation would look like.   
Since
fit parameter errors scale as the square root of the scale of the
data covariance matrix, a factor of $3$ improvement in data errors
has the potential to produce individual CIPT and FOPT fits 
with errors on $\alpha_s$ competitive with those of current 
lattice determinations. Such errors would then be significantly smaller 
than the difference between current CIPT and FOPT results. Theoretical 
progress on the reliability of various perturbative resummation
schemes, as embodied in the current  discrepancy between CIPT and FOPT, will thus 
most likely also be necessary. Whether the outcome of such a
BaBar- or Belle-based analysis will be a more precise determination of 
the strong coupling near the $\t$ mass, or an indication of the
need to construct more sophisticated representations of non-perturbative
effects remains to be seen. Either way, we believe that much can be 
learned from an analysis of the already-existing BaBar and Belle data.

\vspace{3ex}
%\newpage
\noindent {\bf Acknowledgments}
\vspace{3ex}

We would like to thank Martin Beneke, Claude Bernard, Andreas H\"ocker, 
Manel Martinez, and Ramon
Miquel for useful discussions.  We also like to thank
Swagato Banerjee and Sven Menke for significant help with understanding the
HFAG analysis of $\t$ branching fractions, and OPAL spectral-function data, respectively.
MJ and KM thank the 
Department of Physics and Astronomy at SFSU for hospitality.
DB is
supported by the Alexander von Humboldt Foundation, and 
MG is supported in part by the US Department of Energy.
MJ and SP are supported by CICYTFEDER-FPA2008-01430, FPA2011-25948, SGR2009-894,
the Spanish Consolider-Ingenio 2010 Program
CPAN (CSD2007-00042) and SP also by the Programa de Movilidad
PR2010-0284.  
A.M. was supported in part by NASA through Chandra award No.~AR0-11016A, issued by the Chandra X-ray Observatory Center, which is operated by the Smithsonian Astrophysical Observatory for and on behalf of NASA under contract NAS8-03060.
KM is supported by a grant from the Natural Sciences and
Engineering Research Council of Canada. 

%\newpage
\appendix
\section{\label{OPAL} A partial update of the OPAL spectral functions and covariance matrices}
%%####%%
OPAL has made publicly available the spectral functions and covariances
for the three main exclusive modes and inclusive sum
over all modes in each of the $V$ and $A$ channels. The contributions to
the spectral functions corresponding to other exclusive modes (which, with the
exception of $\omega\pi^-\pi^0$, are not measured but constructed
using Monte Carlo) are not available. The covariances between
contributions from different modes are similarly unavailable. 
This limits the extent to which
the OPAL inclusive distributions can be updated for improvements to
the exclusive branching fractions and quantities such as $V_{ud}$
and $B_e$ which enter the conversion between the inclusive differential
decay distributions $dB_{V/A}/ds$ and the spectral functions $\rho_{V/A}(s)$.

The procedure for updating $\rho_{V/A}(s)$ was discussed already in
the text. The ingredients needed for this update are the HFAG
branching fractions and the following $\omega$ and $\eta$ branching
fractions, taken from the 2010 PDG compilation:
\begin{eqnarray}
\label{bfranctions}
B[\omega\rightarrow 3\pi]&=&0.892\pm 0.007\ ,\\
B[\omega\rightarrow\pi^+\pi^- ] &=& 0.0153^{+0.0011}_{-0.0013}\ ,\nonumber\\
B[\eta\rightarrow\pi^+\pi^-\pi^0] &=& 0.2274\pm 0.0028\ .\nonumber
\end{eqnarray}
The latter are needed to convert from the quoted HFAG $3\pi$, $4\pi$ 
and $5\pi$ branching fractions (corresponding to modes defined such that
$\omega$ and $\eta$ substate contributions are absent) to the analogous branching 
fractions of those exclusive modes tabulated by OPAL (defined such that
$\omega$ and $\eta$ substate contributions are included). The corrections 
to be applied to the HFAG branching fractions in order to accomplish this 
conversion include, in addition to those corresponding to the wrong-current 
contaminations discussed already in the main text, 
those corresponding to the contributions of $\omega\pi^-$ 
to the $\pi^-\pi^+\pi^-\pi^0$ distribution and $\omega\pi^0\pi^0$
to the $\pi^-\pi^+\pi^- 2\pi^0$ distribution produced by the 
$\omega\rightarrow\pi^+\pi^-\pi^0$ decay mode. The remainder of
the $\omega\pi^-$ contribution
represents a mode contribution to be assigned to
the $V$ distribution, and likewise, the
$\omega\pi^0\pi^0$ contributions,
and the $\eta\pi^-\pi^0$ (excluding $\eta\rightarrow\pi^+\pi^-\pi^0$)
contribution, represent mode contributions to be 
assigned to the $A$ and residual $V$ distributions in 
the OPAL convention, respectively.
The remainder of the residual mode contributions consist of the
wrong-current contamination corrections and (i) for the $V$ channel,
the $\bar{K}K$, $6\pi$, $\bar{K}K\pi$ and $\bar{K}K\pi\pi$
contributions, and (ii) for the $A$ channel, the $3\pi^- 2\pi^+$,
$\pi^- 4\pi^0$, $\bar{K}K\pi$, $\bar{K}K\pi\pi$ and $a_1$ 
($\rightarrow \pi^-\gamma$) contributions. We follow OPAL in assuming
a fully anti-correlated $50\pm 50\%$ breakdown of the 
$\bar{K}K\pi$ distribution into $V$ and $A$ channel contributions,
and employ the same assumption for the very small $\bar{K}K 2\pi$
contributions not listed by OPAL. The HFAG branching fraction
for the similarly small $a_1$ ($\rightarrow \pi^-\gamma$) mode,
also not listed by OPAL, has also been included in the combined
$A$ residual branching fraction sum.

The inaccessibility of cross-correlations between different
exclusive modes limits our ability to update the OPAL covariance
matrices. We can, however, perform a partial update to take into
account improvements in the determinations of the constant
factors $B_e$, $S_{EW}$ and $V_{ud}$ appearing in the conversion
step
\begin{equation}
\label{rhoB}
\rho_{V/A}(s_k)\, =\, {\frac{dB_{V/A}(s_k)/ds}{B_k}}\ ,
\end{equation}
where $s_k$ is the midpoint of the $k$-th OPAL bin, 
\begin{equation}
\label{Bk}
B_k\, =\, 12\pi^2 S_{EW} \vert V_{ud}\vert^2 B_e\, w_\tau(y_k)/m_\tau^2
\equiv B\, w_\tau (y_k)/m_\tau^2\ ,
\end{equation}
with $y_k=s_k/m_\tau^2$ and $w_\tau (y)$ the $(1+0)$ kinematic
weight $w_\tau (y)=(1-y)^2(1+2y)$. From this it follows that
the relation between the covariances of the spectral function
obtained from the same $dB_{V/A}(s)/ds$ distribution using new (primed) 
and old (unprimed) OPAL values for the constants $S_{EW}$, $V_{ud}$,
$B_e$ and $m_\tau^2$, incorporating also, for completeness, in the
updated version, the contributions of the
uncertainty on $m_\tau$ neglected by OPAL, is ($\r_i$ runs over all
$\r_V(s_i)$ and $\r_A(s_i)$)
\begin{eqnarray}
\label{covupdate}
\langle \delta\rho^\prime_i\delta\rho^\prime_j\rangle \, &=&\,
{\frac{B_i B_j}{B^\prime_i B^\prime_j}}\left[
\langle \delta\rho_i \delta \rho_j\rangle \, +\, \rho_i \rho_j\,
\left( \left({\frac{ \delta B^\prime}{B^\prime}}\right)^2\, -\,
\left({\frac{ \delta B}{B}}\right)^2\right) \right. \\
&&\left. \ \ +\, \left({\frac{\delta m_\tau}{m_\tau}}\right)^2\,
\left( {\frac{(-2+18y_i^2-16y_i^3)(-2+18y_j^2-16y_j^3)}{w_\tau (y_i)
w_\tau (y_j)}}\right)\, \rho_i\rho_j\right] .
\nonumber
\end{eqnarray}
For the current values of the physical quantities appearing in these conversions,
we will use
\begin{eqnarray}
\label{constants}
S_{EW}&=&1.0201(3)\ ,\\
|V_{ud}|&=&0.97425(22)\ ,\nonumber\\
B_e&=&0.17827(40)\ ,\nonumber\\
m_\t&=&1.77677(15)~\mbox{GeV}\ ,\nonumber
\end{eqnarray}
from Refs.~\cite{erlersew,htrpp10} and \cite{HFAG} for $B_e$ and $m_\t$, respectively.   The error on
$m_\t$ plays no significant role in our analysis.

%%%%%%%%%%%%%%%%%%%%%%%%%%%%%%%%%%%%%%%%%%%%%%%%%%%%%%%%%%%%%%%%%%%%%%%%%
%\newpage

\end{document}